\documentclass[aps,prb,10pt,twocolumn,superscriptaddress]{revtex4-2}

\usepackage{tikz,pgf}
\usepackage{amsfonts}
\usepackage{amsmath}
\usepackage{amssymb}
\usepackage{mathtools}
\usepackage{braket}
\usepackage[right]{lineno}
\usepackage{algorithmicx,algpseudocode}
\usepackage[linesnumbered,ruled,vlined]{algorithm2e}
\usepackage{dcolumn}
\usepackage{listings}
\usepackage{subcaption}
\usepackage{caption}
\usepackage{enumitem}
\usepackage{bbm}
\usepackage{orcidlink}
\usepackage{graphicx}
\usepackage{xcolor}

\newcolumntype{d}[1]{D{.}{.}{#1}}

\graphicspath{{figures/}}
\makeatletter
\def\input@path{{diagrams/}}
\makeatother

\definecolor{darkgreen}{rgb}{0.0, 0.5, 0.0} 
\definecolor{codegreen}{rgb}{0,0.6,0}
\definecolor{codegray}{rgb}{0.5,0.5,0.5}
\definecolor{codepurple}{rgb}{0.58,0,0.82}
\definecolor{backcolour}{rgb}{0.95,0.95,0.92}

\lstdefinestyle{mystyle}{
    backgroundcolor=\color{backcolour},   
    commentstyle=\color{codegreen},
    keywordstyle=\color{magenta},
    numberstyle=\tiny\color{codegray},
    stringstyle=\color{codepurple},
    basicstyle=\ttfamily\footnotesize,
    breakatwhitespace=false,         
    breaklines=true,                 
    captionpos=b,                    
    keepspaces=true,                 
    numbers=left,                    
    numbersep=5pt,                  
    showspaces=false,                
    showstringspaces=false,
    showtabs=false,                  
    tabsize=2
}
\usetikzlibrary{plotmarks, intersections, calc, positioning, arrows, shapes.geometric, matrix, decorations.markings, decorations.pathreplacing}

\tikzset{
    every path/.style={thick}, 
    >=stealth
}

\DeclareMathOperator{\rank}{rank}

\newcommand{\C}{\mathbb{C}}

\newcommand{\Z}{\mathbb{Z}}
\newcommand{\Q}{\mathbb{Q}}
\newcommand{\R}{\mathbb{R}}
\newcommand{\id}{\mathbbm{1}}

\def\kbbox[#1,#2,#3,#4,#5]#6{
        \draw[dashed] node[draw,color=gray!50,minimum
        height=#1,minimum width=#2] (#4) at #5 {}; 
        \node[anchor=#3,inner sep=2pt] at (#4.#3)  {#6};
        }

\begin{document}

\title{Optimal Symbolic Construction of Matrix Product Operators and Tree Tensor Network Operators}

\author{Hazar Çakır \orcidlink{0009-0005-3271-9110}}
\email{hazar.cakir@tum.de}
\affiliation{Technical University of Munich, CIT, Department of Computer Science, Boltzmannstra{\ss}e 3, 85748 Garching, Germany}
\author{Richard M.~Milbradt \orcidlink{0000-0001-8630-9356}}
\email{r.milbradt@tum.de}
\affiliation{Technical University of Munich, CIT, Department of Computer Science, Boltzmannstra{\ss}e 3, 85748 Garching, Germany}
\author{Christian B.~Mendl \orcidlink{0000-0002-6386-0230}}
\email{christian.mendl@tum.de}
\affiliation{Technical University of Munich, CIT, Department of Computer Science, Boltzmannstra{\ss}e 3, 85748 Garching, Germany}
\affiliation{Technical University of Munich, Institute for Advanced Study, Lichtenbergstra{\ss}e 2a, 85748 Garching, Germany}

\begin{abstract}
This research introduces an improved framework for constructing matrix product operators (MPOs) and tree tensor network operators (TTNOs), crucial tools in quantum simulations. A given (Hamiltonian) operator typically has a known symbolic ``sum of operator strings'' form that can be translated into a tensor network structure. Combining the existing bipartite-graph-based approach and a newly introduced symbolic Gaussian elimination preprocessing step, our proposed method improves upon earlier algorithms in cases when Hamiltonian terms share the same prefactors. We test the performance of our method against established ones for benchmarking purposes. Finally, we apply our methodology to the model of a cavity filled with molecules in a solvent. This open quantum system is cast into the hierarchical equation of motion (HEOM) setting to obtain an effective Hamiltonian. Construction of the corresponding TTNO demonstrates a sub-linear increase of the maximum bond dimension.
\end{abstract}

\maketitle

\section{Introduction}
\label{sec:intro}

Simulating quantum systems on classical computers presents an inherent challenge: the representation of quantum systems. Due to the fundamentally different nature of quantum data compared to classical approaches, the resource requirements to represent a quantum system on a classical computer generally scale exponentially with the system's size. Over time, methods have been developed to alleviate this problem. A cornerstone of these efforts is tensor network methods \cite{Schollwock2011, Bridgeman_2017}, which provide a suitable framework to represent quantum states \cite{PhysRevLett.93.207204} and quantum operators \cite{Pirvu_2010}. When the quantum system's Hamiltonian is written as matrix product operator (MPO), one can utilize the well-known DMRG algorithm for ground state search \cite{Schollwock2011} or the time-dependent variational principle for time evolution (TDVP) \cite{Lubich2015, Haegeman2016}. There has been research into the construction and optimization of MPOs \cite{Pirvu_2010, PhysRevB.95.035129, McCulloch2007, 10.1063/1.4939000, Frowis2010, Ren_2020}. However, these methods cannot handle symbolic coefficient representations, can only be applied to a specific class of quantum systems, or might not find the optimal MPO in all cases. We aim to address this gap by developing an algorithm that starts with a quantum Hamiltonian's analytic and symbolic description and generates an optimal MPO. By optimal, we mean achieving the lowest possible virtual bond dimensions in the resulting tensor network. This is essential, as the computational cost of tensor network algorithms typically scales polynomially with the bond dimension. Small bond dimensions directly reduce memory usage and computational time, enabling the simulation of larger or more complex quantum systems. Importantly, our approach does not perform an approximate compression but instead constructs an exact representation of the original Hamiltonian, preserving all mathematical properties.

Another key structure explored in this work is a tree tensor network operator (TTNO) \cite{Szalay2015, Frowis2010, 10.1063/5.0218773}, an alternative representation of quantum systems based on a tree structure for the tensor network \cite{PhysRevA.74.022320_ttno}. We provide a formal introduction to TTNOs, along with the theoretical background of tensor networks and operator construction, in Section~\ref{sec:theory}. Unlike MPOs, which impose a linear representation regardless of the system's inherent structure, TTNOs offer a framework better suited for capturing long-range interactions by maintaining a tree-like relationship scheme, which is more accurate for certain scenarios \cite{Cirac2021}. TTNOs are used in adapted versions of DMRG \cite{Nakatani_2013} and TDVP \cite{Bauernfeind2020} as well as algorithms that make explicit use of the tree structure, such as the basis-update and Galerkin method (BUG) \cite{Ceruti2021, Ceruti2024}.

A basis for our work is the application of bipartite graph theory for constructing MPOs and TTNOs as proposed in \cite{Ren_2020, 10.1063/5.0218773}. Bipartite graph theory provides an efficient framework for the symbolic minimization of the virtual bond dimensions. However, this algorithm does not cover all possible cases, failing to find optimal tensor network operators when the terms in a Hamiltonian are interdependent (e.g., share the same coefficients). Our proposed approach incorporates symbolic Gaussian elimination as a preprocessing step for the bipartite graph algorithm, which is explained in detail in Section~\ref{sec:methodology}. This enhancement preserves the symbolic nature of the original solution while introducing improvements to tackle the limitations. Subsequently, we adapt the algorithm for TTNOs. This adaptation presented significant challenges for an efficient implementation. While the core principles of bond optimization remain unchanged, considerable effort was required at the implementation level to accommodate tree structures. We test and compare the performance of our proposed algorithm against the original bipartite graph algorithm in Section~\ref{sec:results}. Finally, in Section~\ref{sec:conclusions}, we discuss the implications of our findings and outline potential directions for future research.

\begin{figure*}[ht]
    \centering
    \begin{subfigure}[b]{0.3\textwidth}
        \centering
        \begin{tikzpicture}[default node/.style={draw, circle, minimum size=0.8cm, thick}]
\def\vertdist{1}
\def\horizdist{1}

\node[default node,draw] (N1) at (0,0) {$1$};
\node[default node, draw] (N2) at (-\horizdist,-\vertdist) {$2$};
\node[default node, draw] (N3) at (\horizdist,-\vertdist) {$3$};
\node[default node, draw] (N4) at (0*\horizdist,-2*\vertdist) {$4$};
\node[default node, draw] (N5) at (2*\horizdist,-2*\vertdist) {$5$};

\draw (N2) -- (N1) -- (N3) -- (N4);
\draw (N3) -- (N5);
\end{tikzpicture}
        \caption{TTN without physical legs}
        \label{fig:tree_structure}
    \end{subfigure}
    \begin{subfigure}[b]{0.3\textwidth}
        \centering
        \begin{tikzpicture}[default node/.style={draw, circle, minimum size=0.8cm, inner sep=0, thick}]
\def\vertdist{1}
\def\horizdist{1}
\def\physdist{0.75}

\node[default node,draw] (N1) at (0,0) {$T^{(1)}$};
\node[default node, draw] (N2) at (-\horizdist,-\vertdist) {$T^{(2)}$};
\node[default node, draw] (N3) at (\horizdist,-\vertdist) {$T^{(3)}$};
\node[default node, draw] (N4) at (0*\horizdist,-2*\vertdist) {$T^{(4)}$};
\node[default node, draw] (N5) at (2*\horizdist,-2*\vertdist) {$T^{(5)}$};

\draw (N2) -- (N1) -- (N3) -- (N4);
\draw (N3) -- (N5);

\draw[very thick] (N1) -- (0,\physdist);
\draw[very thick] (N2) -- (-\horizdist,-\vertdist+\physdist);
\draw[very thick] (N3) -- (\horizdist,-\vertdist+\physdist);
\draw[very thick] (N4) -- (0*\horizdist,-2*\vertdist+\physdist);
\draw[very thick] (N5) -- (2*\horizdist,-2*\vertdist+\physdist);

\end{tikzpicture}

        \caption{TTNS}
        \label{fig:ttns}
    \end{subfigure}
    \begin{subfigure}[b]{0.3\textwidth}
        \centering
        \begin{tikzpicture}[default node/.style={draw, circle, minimum size=0.8cm, inner sep=0, thick}]
\def\vertdist{1}
\def\horizdist{1}
\def\physdist{0.75}

\node[default node,draw] (N1) at (0,0) {$\mathcal{T}^{(1)}$};
\node[default node, draw] (N2) at (-\horizdist,-\vertdist) {$\mathcal{T}^{(2)}$};
\node[default node, draw] (N3) at (\horizdist,-\vertdist) {$\mathcal{T}^{(3)}$};
\node[default node, draw] (N4) at (0*\horizdist,-2*\vertdist) {$\mathcal{T}^{(4)}$};
\node[default node, draw] (N5) at (2*\horizdist,-2*\vertdist) {$\mathcal{T}^{(5)}$};

\draw (N2) -- (N1) -- (N3) -- (N4);
\draw (N3) -- (N5);

\draw[very thick] (N1) -- (0,\physdist);
\draw[very thick] (N2) -- (-\horizdist,-\vertdist+\physdist);
\draw[very thick] (N3) -- (\horizdist,-\vertdist+\physdist);
\draw[very thick] (N4) -- (0*\horizdist,-2*\vertdist+\physdist);
\draw[very thick] (N5) -- (2*\horizdist,-2*\vertdist+\physdist);
\draw[very thick] (N1) -- (0,-\physdist);
\draw[very thick] (N2) -- (-\horizdist,-\vertdist-\physdist);
\draw[very thick] (N3) -- (\horizdist,-\vertdist-\physdist);
\draw[very thick] (N4) -- (0*\horizdist,-2*\vertdist-\physdist);
\draw[very thick] (N5) -- (2*\horizdist,-2*\vertdist-\physdist);

\end{tikzpicture}
        \caption{TTNO}
        \label{fig:ttno}
    \end{subfigure}
    \caption{Examples of tensor networks with the same tree topology, differing in the physical dimensions.}
    \label{fig:trees}
\end{figure*}
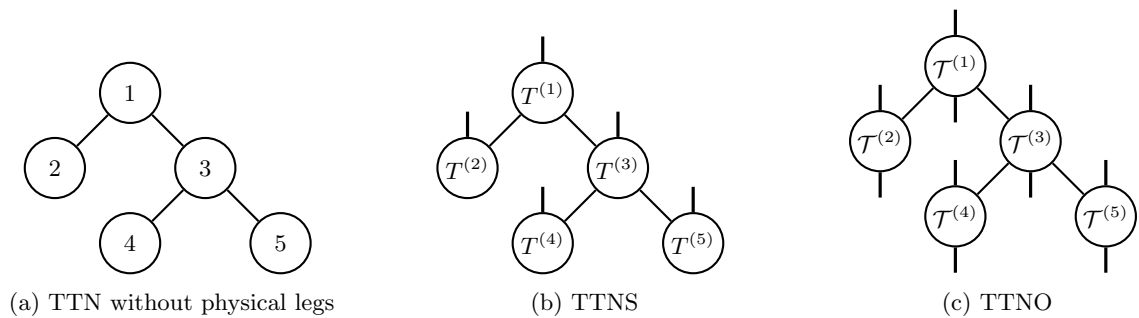

The code written for the methods in this work is based on the PyTreeNet library \cite{milbradt2024pytreenet} and can be found at \cite{PyTreeNet}. The dataset used for the experiments and evaluations in this work is available at \cite{1773144} and includes all generated data and plots. 

\section{Theory}
\label{sec:theory}

The reviews \cite{Schollwock2011, Bridgeman_2017} provide a didactical introduction to tensor network methods in general. In the following, we focus on the mathematical framework of tree tensor networks and state diagrams for operator construction.

\subsection{Tree Tensor Networks}
Tree tensor networks (TTNs) are a subclass of tensor networks with a tree topology connectivity \cite{PhysRevA.74.022320_ttno, Murg_2010}. The hierarchical nature of TTNs ensures a balanced distribution of information across the network and allows for efficient representation and manipulation of quantum states. They can approximate more highly entangled states than MPSs \cite{Nakatani_2013, Gunst2018} while avoiding difficulties due to loops like in projected entangled-pair states (PEPS).

Like any other tensor network, a TTN is represented by a set of tensors \( \left\{T^{(1)}, T^{(2)}, \ldots, T^{(L)}\right\} \), where each tensor \( T^{(\ell)} \in \C^{d_1 \times \cdots \times d_{m_\ell}} \). The tensor legs represent these dimensions. Legs with equal dimensions connected in a tree structure are called virtual legs (or bonds). Examples of TTNs are given in Fig.~\ref{fig:trees}, here the TTN in Fig.~\ref{fig:tree_structure} corresponds to a scalar.

Leaving some legs in a TTN unconnected allows us to represent quantum states. In this case, the unconnected legs represent the physical dimension of the state and are fittingly referred to as physical legs. Assuming every tensor has one physical leg the state is obtained from the TTN via
\begin{equation}\label{eq:ttns}
\begin{split}
\ket{\psi} = \sum_{\substack{p_1, \ldots, p_L\\
                            v_1, \ldots, v_M}} &  T^{(1)}_{\{v_{1,i}\}, p_1} \ldots T^{(L)}_{\{v_{L,i}\}, p_L} \\
                            & \cdot \ket{p_1, p_2, \ldots, p_L },
\end{split}
\end{equation}
where $L$ is the number of physical sites, $M$ is the overall number of virtual legs, and $\{v_{\ell,i}\} \subset \{v_1, \ldots, v_M\}$ and $p_\ell$ are the indices of the virtual and physical legs of the tensor $T^{(\ell)}$ respectively. $\ket{\psi}$ or rather the tensor network consisting of $\left\{T^{(1)}, T^{(2)}, \ldots, T^{(L)}\right\}$ in Eq.~\eqref{eq:ttns} is called tree tensor network state (TTNS). The example of a TTNS in Fig.~\ref{fig:ttns} represents a quantum state with five physical legs. Note that some physical legs can serve as a dummy index with a trivial dimension of $1$.

\subsection{Tree Tensor Network Operators}

The concept of TTNs can be extended to tree tensor network operators (TTNOs). These are particularly useful for representing and manipulating many-body operators, such as Hamiltonians, in a structured and efficient manner. The construction of a TTNO involves decomposing a global operator into a hierarchical set of local operators and intermediate tensors, similar to how a TTNS decomposes a global state.

We denote a TTNO as a set of operator tensors \( \{\mathcal{T}_1, \mathcal{T}_2, \ldots, \mathcal{T}_L\} \), where each tensor has exactly two physical legs. We can then obtain the corresponding operator via
\begin{equation}\label{eq:ttno}
\begin{split}
    \hat{O} = \sum_{\substack{p_1, \ldots, p_L\\
                              q_1, \ldots, q_L\\
                              v_1, \ldots, v_M}} &
                    \mathcal{T}^{(1)}_{\{v_{1,i}\}, p_1, q_1} \cdots \mathcal{T}^{(L)}_{\{v_{L,i}\}, p_L, q_L} \\
                    & \cdot \ket{p_1, p_2, \ldots, p_L}\bra{q_1, q_2, \ldots, q_L},
\end{split}
\end{equation}
where $\{v_i\}_\ell \subset \{v_1, \ldots, v_M\}$ are the virtual leg indices and $p_\ell$ and $q_\ell$ are the indices of the physical legs of the tensor $\mathcal{T}^{(\ell)}$. Fig.~\ref{fig:ttno} shows an example of a TTNO. Here, each tensor has two physical legs, which could be trivial dummy legs.

\subsection{Hamiltonian Representation with Operator Strings}

We consider many-body operators, with a focus on Hamiltonians, of the form
\begin{equation}
\label{eq:sum_prod_op_repr}
\hat{O} = \sum_{k=1}^K \hat{O}_k = \sum_{k=1}^K \gamma_k \, \hat{O}_k^{(1)} \otimes \hat{O}_k^{(2)} \otimes \cdots \otimes \hat{O}_k^{(L)},
\end{equation}
where \( \gamma_k \) are real or complex coefficients associated with each term, and \( \hat{O}_k^{(\ell)} \) represent local operators acting on different sites \( \ell \in \{ 1, \dots, L \} \) of the system. The coefficients \( \gamma_k \) determine the system's physical properties and represent the strength and nature of the interactions encoded in the Hamiltonian. These coefficients might correspond to coupling constants, external field strengths, or other system dynamics parameters in physical models.

This sum of product representation offers several advantages when working with complex quantum systems. First, it provides \emph{modularity}, where each term in the Hamiltonian is represented as a modular unit comprising a coefficient and a sequence of local operators. This modularity facilitates the systematic construction and manipulation of the Hamiltonian, making it easier to introduce or modify new interactions. Second, it is very \emph{general}. Any operator in a product Hilbert space can be written in the form \eqref{eq:sum_prod_op_repr} via local operator bases, and many relevant operators are defined as such a linear combination \cite{chan2016matrix, Paeckel_2019} or can be approximated in this way \cite{Sasmal_2024}. Lastly, each term in the Hamiltonian can be directly mapped onto a desired tensor network structure, with the local operators corresponding to tensors and the coefficients \( \gamma_k \) incorporated as weights. This makes it highly \emph{compatible} with MPOs and TTNOs.

\subsection{State Diagrams for TTNOs and MPOs}

\begin{figure}[t]
    \centering
    \begin{tikzpicture}[
default node/.style={draw, minimum size=0.5cm, thick},
edge node/.style={fill, circle, scale=0.5}
]
\def\labeldist{1}
\def\vertdist{1.2}
\def\distnodehe{\vertdist}

\node[default node,draw] (X1) at (0,0) {$X^{(1)}$};
\node[default node,draw] (Y1) at (\labeldist,0) {$Y^{(1)}$};
\node[default node,draw] (I1) at (2*\labeldist,0) {$\id^{(1)}$};

\begin{scope}[shift={(-1.5,-\vertdist)}]
    \node[edge node] (EN121) at (0,0){};
    \node[edge node] (EN122) at (\labeldist,0){};

    \node[default node,draw] (X2) at (0,-\distnodehe) {$X^{(2)}$};
    \node[default node,draw] (I2) at (\labeldist,-\distnodehe) {$\id^{(2)}$};
\end{scope}

\begin{scope}[shift={(1.5,-\vertdist)}]
    \node[edge node] (EN131) at (0,0){};
    \node[edge node] (EN132) at (\labeldist,0){};

    \node[default node,draw] (I3) at (0,-\distnodehe) {$\id^{(3)}$};
    \node[default node,draw] (Y3) at (\labeldist,-\distnodehe) {$Y^{(3)}$};
    \node[default node,draw] (Z3) at (2*\labeldist,-\distnodehe) {$Z^{(3)}$};
\end{scope}

\begin{scope}[shift={(1,-3.2*\vertdist)}]
    \node[edge node] (EN341) at (\labeldist,0){};
    \node[edge node] (EN342) at (0,0){};

    \node[default node,draw] (Y4) at (0,-\distnodehe/2) {$Y^{(4)}$};
    \node[default node,draw] (I4) at (\labeldist,-\distnodehe/2) {$\id^{(4)}$};
\end{scope}

\begin{scope}[shift={(4,-3.2*\vertdist)}]
    \node[edge node] (EN351) at (0,0){};
    \node[default node,draw] (I5) at (0,-\distnodehe/2) {$\id^{(5)}$};
\end{scope}

\draw (X2) -- (EN121) to[out=90,in=-90] (X1);
\draw (I2) -- (EN122) to[out=90,in=-90] (Y1);
\draw (EN122) to[out=90,in=-90,looseness=0.5] (I1);

\draw (X1) to[out=-90,in=90] (EN131) -- (I3);
\draw (I1) to[out=-90,in=90] (EN131) to[out=-90,in=90] (Z3);
\draw (Y1) to[out=-90,in=90] (EN132) -- (Y3);

\draw (I3) to[out=-90,in=90] (EN342) -- (Y4);
\draw (Y3) to[out=-90,in=90] (EN342);
\draw (Z3) to[out=-90,in=90] (EN341) -- (I4);

\draw (I3) to[out=-90,in=90] (EN351) -- (I5);
\draw (Y3) to[out=-90,in=90] (EN351);
\draw (Z3) to[out=-90,in=90] (EN351);

\end{tikzpicture}

    \caption{An example state diagram representing a TTNO with the structure given in Fig.~\ref{fig:ttno}. The black dots correspond to the vertices $\{\nu_i\}$, and the boxes to the hyperedges $\{e_i\}$. The corresponding operator is $\hat{O}=X^{(1)}X^{(2)}Y^{(4)}+X^{(1)}X^{(2)}Z^{(3)}$ $+Y^{(1)}Y^{(3)}Y^{(4)}+Y^{(4)}+Z^{(3)}$.}
    \label{fig:state_diagram}
\end{figure}
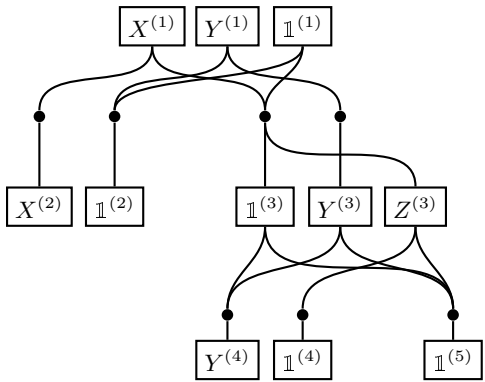

\begin{figure*}[ht]
    \centering
    \begin{tikzpicture}[thick, 
                    site/.style={draw, minimum size=0.7cm,font=\large},
                    connection/.style={draw, solid, thick},
                    blackdot/.style={fill=black, circle, minimum size=5pt, inner sep=0pt},
                    default node/.style={draw, circle, minimum size=1cm, font=\large},
    scale=0.68, transform shape
    ]
        \begin{scope}[shift={(0,0)}]
            \node[site] (site11) at (0,0) {$X^{(1)}$};
            \node[blackdot] (dot12) at (1,0) {};
            \node[site] (site12) at (2,0) {$\id^{(2)}$};
            \node[blackdot] (dot13) at (3,0) {};
            \node[site] (site13) at (4,0) {$Y^{(3)}$};
            \node[blackdot] (dot14) at (5,0) {};
            \node[site] (site14) at (6,0) {$Y^{(4)}$};
            \node[blackdot] (dot15) at (7,0) {};
            \node[site] (site15) at (8,0) {$\id^{(5)}$};
            
            \node[site] (site21) at (0,-1.5) {$X^{(1)}$};
            \node[blackdot] (dot22) at (1,-1.5) {};
            \node[site] (site22) at (2,-1.5) {$\id^{(2)}$};
            \node[blackdot] (dot23) at (3,-1.5) {};
            \node[site] (site23) at (4,-1.5) {$Z^{(3)}$};
            \node[blackdot] (dot24) at (5,-1.5) {};
            \node[site] (site24) at (6,-1.5) {$\id^{(4)}$};
            \node[blackdot] (dot25) at (7,-1.5) {};
            \node[site] (site25) at (8,-1.5) {$X^{(5)}$};
            
            \node[site] (site31) at (0,-3) {$\id^{(1)}$};
            \node[blackdot] (dot32) at (1,-3) {};
            \node[site] (site32) at (2,-3) {$\id^{(2)}$};
            \node[blackdot] (dot33) at (3,-3) {};
            \node[site] (site33) at (4,-3) {$X^{(3)}$};
            \node[blackdot] (dot34) at (5,-3) {};
            \node[site] (site34) at (6,-3) {$X^{(4)}$};
            \node[blackdot] (dot35) at (7,-3) {};
            \node[site] (site35) at (8,-3) {$\id^{(5)}$};
        
            \node[site] (site41) at (0,-4.5) {$Z^{(1)}$};
            \node[blackdot] (dot42) at (1,-4.5) {};
            \node[site] (site42) at (2,-4.5) {$\id^{(2)}$};
            \node[blackdot] (dot43) at (3,-4.5) {};
            \node[site] (site43) at (4,-4.5) {$X^{(3)}$};
            \node[blackdot] (dot44) at (5,-4.5) {};
            \node[site] (site44) at (6,-4.5) {$X^{(4)}$};
            \node[blackdot] (dot45) at (7,-4.5) {};
            \node[site] (site45) at (8,-4.5) {$\id^{(5)}$};
        
            \draw[connection] (site11) -- (dot12);
            \draw[connection] (dot12) -- (site12);
            \draw[connection] (site12) -- (dot13);
            \draw[connection] (dot13) -- (site13);
            \draw[connection] (site13) -- (dot14);
            \draw[connection] (dot14) -- (site14);
            \draw[connection] (site14) -- (dot15);
            \draw[connection] (dot15) -- (site15);
            
            \draw[connection] (site21) -- (dot22);
            \draw[connection] (dot22) -- (site22);
            \draw[connection] (site22) -- (dot23);
            \draw[connection] (dot23) -- (site23);
            \draw[connection] (site23) -- (dot24);
            \draw[connection] (dot24) -- (site24);
            \draw[connection] (site24) -- (dot25);
            \draw[connection] (dot25) -- (site25);
            
            \draw[connection] (site31) -- (dot32);
            \draw[connection] (dot32) -- (site32);
            \draw[connection] (site32) -- (dot33);
            \draw[connection] (dot33) -- (site33);
            \draw[connection] (site33) -- (dot34);
            \draw[connection] (dot34) -- (site34);
            \draw[connection] (site34) -- (dot35);
            \draw[connection] (dot35) -- (site35);
        
            \draw[connection] (site41) -- (dot42);
            \draw[connection] (dot42) -- (site42);
            \draw[connection] (site42) -- (dot43);
            \draw[connection] (dot43) -- (site43);
            \draw[connection] (site43) -- (dot44);
            \draw[connection] (dot44) -- (site44);
            \draw[connection] (site44) -- (dot45);
            \draw[connection] (dot45) -- (site45);
        \end{scope}
    
            \draw[->, thick] (8.8, -2) -- +(1,0);
    
        \begin{scope}[shift={(10.4,0)}]
            \node[site] (site11) at (0,-0.75) {$X^{(1)}$};
            \node[blackdot] (dot12) at (1,-0.75) {};
            \node[site] (site12) at (2,-0.75) {$\id^{(2)}$};
            \node[blackdot] (dot13) at (3,-0.75) {};
            \node[site] (site13) at (4,0) {$Y^{(3)}$};
            \node[blackdot] (dot14) at (5,0) {};
            \node[site] (site14) at (6,0) {$Y^{(4)}$};
            \node[blackdot] (dot15) at (7,0) {};
            \node[site] (site15) at (8,0) {$\id^{(5)}$};
            
            \node[site] (site23) at (4,-1.5) {$Z^{(3)}$};
            \node[blackdot] (dot24) at (5,-1.5) {};
            \node[site] (site24) at (6,-1.5) {$\id^{(4)}$};
            \node[blackdot] (dot25) at (7,-1.5) {};
            \node[site] (site25) at (8,-1.5) {$X^{(5)}$};
            
            \node[site] (site31) at (0,-3) {$\id^{(1)}$};
            \node[blackdot] (dot32) at (1,-3) {};
            \node[site] (site32) at (2,-3) {$\id^{(2)}$};
            \node[blackdot] (dot33) at (3,-3.75) {};
            \node[site] (site33) at (4,-3.75) {$X^{(3)}$};
            \node[blackdot] (dot34) at (5,-3.75) {};
            \node[site] (site34) at (6,-3.75) {$X^{(4)}$};
            \node[blackdot] (dot35) at (7,-3.75) {};
            \node[site] (site35) at (8,-3.75) {$\id^{(5)}$};
        
            \node[site] (site41) at (0,-4.5) {$Z^{(1)}$};
            \node[blackdot] (dot42) at (1,-4.5) {};
            \node[site] (site42) at (2,-4.5) {$\id^{(2)}$};

            \draw[connection] (site11) -- (dot12);
            \draw[connection] (dot12) -- (site12);
            \draw[connection] (site12) -- (dot13);
            \draw[connection] (dot13) -- (site13);
            \draw[connection] (dot13) -- (site23);
            \draw[connection] (site13) -- (dot14);
            \draw[connection] (dot14) -- (site14);
            \draw[connection] (site14) -- (dot15);
            \draw[connection] (dot15) -- (site15);
            
            \draw[connection] (site23) -- (dot24);
            \draw[connection] (dot24) -- (site24);
            \draw[connection] (site24) -- (dot25);
            \draw[connection] (dot25) -- (site25);
            
            \draw[connection] (site31) -- (dot32);
            \draw[connection] (dot32) -- (site32);
            \draw[connection] (site32) -- (dot33);
            \draw[connection] (dot33) -- (site33);
            \draw[connection] (site33) -- (dot34);
            \draw[connection] (dot34) -- (site34);
            \draw[connection] (site34) -- (dot35);
            \draw[connection] (dot35) -- (site35);
        
            \draw[connection] (site41) -- (dot42);
            \draw[connection] (dot42) -- (site42);
            \draw[connection] (site42) -- (dot33);

        
            \draw[dashed, color=red, thick] (3, 0.5) -- (3, -5)
            node[at start,yshift=25pt, color=red, font=\large] {Cut site};

            \draw[decorate,decoration={brace,amplitude=5pt},thick, color=orange] (-0.4,0.5) -- (2.4,0.5) node[midway,yshift=16pt, color=orange, font=\large] {$U$: Left chains};
            \draw[decorate,decoration={brace,amplitude=5pt},thick, color=darkgreen] (3.4,0.5) -- (8.4,0.5) node[midway,yshift=16pt, color=darkgreen, font=\large] {$V$: Right chains};
        \end{scope}
        
        \draw[->, thick] (19.2, -2) -- +(1,0);

        \begin{scope}[shift={(21,0)}]
            \def\vdist{3.5cm}
            \def\hdist{4}
            
            \node[draw,circle, default node] (U1) at (0,0) {$U_1$};
            \node[draw,circle, default node] (U2) at (0,-0.5*\vdist) {$U_2$};
            \node[draw,circle, default node] (U3) at (0,-1*\vdist) {$U_3$};
            \node[draw,circle, default node] (V1) at (\hdist,0*\vdist) {$V_1$};
            \node[draw,circle, default node] (V2) at (\hdist,-0.5*\vdist) {$V_2$};
            \node[draw,circle, default node] (V3) at (\hdist,-1*\vdist) {$V_3$};

            \node[draw,circle,fill=black,inner sep=0, minimum size=10] (b1) at (0.5*\hdist,0*\vdist){};
            \node[draw,circle,fill=black,inner sep=0, minimum size=10] (b2) at (0.5*\hdist,-0.75*\vdist){};
           
            \draw (U1) -- (b1) -- (V1) ;
            \draw (b1) to[out=0,in=180] (V2);

            \draw (b2) -- (V3);
            \draw (U2) to[out=0,in=180] (b2);
            \draw (U3) to[out=0,in=180] (b2);
            
        \end{scope}
    \end{tikzpicture}
    \caption{An example state diagram representing an MPO of the operator $\hat{O}=X^{(1)}Y^{(3)}Y^{(4)}+X^{(1)}Z^{(3)}X^{(5)}$ $+X^{(3)}X^{(4)}+Z^{(1)}X^{(3)}X^{(4)}$ on a chain of quantum systems. The leftmost part of the figure shows a collection of unoptimized operator strings. In the middle, a vertical dashed line marks a cut between site 2 and site 3, partitioning the operator chains into left and right halves. These are grouped into sets \( U \) and \( V \), where identical subchains are merged. The rightmost subfigure shows the corresponding bipartite graph: nodes \( U_i \in U \) and \( V_j \in V \) are connected by edges representing terms in the original operator decomposition. 
}
    \label{fig:state_diagram_mpo}
\end{figure*}

State diagrams are graphical representations that capture the relationships between the tensor elements in a network. They show which elements are multiplied together during a tensor contraction \cite{Crosswhite2008}. State diagrams are especially useful when considering TTNs and, more specifically, the construction of TTNOs from operators as given in Eq.~\eqref{eq:sum_prod_op_repr} \cite{milbradt2024state}.

A state diagram \( \mathcal{D} = (V, E) \) consists of vertices and hyperedges and represents a TTNO. The vertices, denoted as \( V = \left\{\nu_1, \nu_2, \ldots, \nu_n\right\} \), connect multiple hyperedges. They represent the tensor contractions between the operator tensors $\mathcal{T}^{(\ell)}$, corresponding to the shared indices (i.e., virtual bonds) in the tensor network. The hyperedges, denoted as \( E = \{e_1, e_2, \ldots, e_m\} \), represent the operator tensors $\mathcal{T}^{(\ell)}$. Each hyperedge has a label corresponding to an operator acting on the site $\ell$ in the quantum system. Thus there is a collection of hyperedges $\mathcal{V}_\ell$ corresponding to site $\ell$ and a collection of vertices $\mathcal{E}_{\ell,\ell'}$ corresponding to the virtual leg connecting sites $\ell$ and $\ell'$. An example optimized state diagram based on the TTNO in Fig.~\ref{fig:ttno} is provided in Fig.~\ref{fig:state_diagram}. The different terms can be obtained by choosing a exactly one hyperedge from each $\mathcal{V}_\ell$ and $\mathcal{E}_{\ell,\ell'}$ and ensuring they are connected.

As MPO are a special case of TTNO, also any MPO can be represented using state diagrams. Figure~\ref{fig:state_diagram_mpo} illustrates an unoptimized MPO structure comprising several example operator strings. The diagram is intended to provide both a representative MPO example and an illustration of how subchains can be processed and grouped—topics that will be discussed in detail in Section~\ref{sec:methodology}. To demonstrate this, we highlight a bipartition of the operator chain between site 2 and site 3. In this context, we refer to the sets of operator subchains to the left and right of the cut as \( U \) and \( V \), respectively. Each node \( U_i \in U \) corresponds to a unique left-hand subchain, while each node \( V_j \in V \) represents a right-hand subchain. These partitions, \( U \) and \( V \), serve as the foundational elements of the bipartite graph construction employed throughout the paper.

State diagrams offer a clear and intuitive way to visualize the complex tensor networks underlying TTNOs. They also facilitate the identification of optimal contraction sequences and the minimization of bond dimensions, making them a valuable tool for theoretical analysis and practical implementation. For a discussion on how quantum numbers and Abelian symmetries can be incorporated into these diagrams, see Appendix~\ref{appendix:quantum_numbers}.

\section{Methodology - Algorithm}
\label{sec:methodology}

Given an operator of the form \eqref{eq:sum_prod_op_repr}, can we automatically construct an MPO or TTNO with minimum bond dimension? Our proposed algorithm to solve this problem first considers the local optimization of a single bond using a modified version of the bipartite graph optimization introduced in \cite{Ren_2020}. We then adapt the method to build TTNOs with an arbitrary underlying tree topology, cf.~\cite{10.1063/5.0218773}.

\subsection{Optimization of a Local Bond Dimension}
\subsubsection{The Bipartite Graph Method}\label{sec:bipartite_method}
The first part of our algorithm focuses on optimizing local bond dimensions, leveraging an existing method based on bipartite graph theory introduced in \cite{Ren_2020}. This algorithm can only find the optimal bond dimension if all prefactors $\gamma_i$ of Eq.~\eqref{eq:sum_prod_op_repr} are different and independent. A minimum example where the assumption is not valid, and the algorithm of \cite{Ren_2020} cannot find the minimal bond dimension is shown in Appendix~\ref{fig:bipartite_fc}.

We briefly recall two key graph-theoretical concepts used in this work: the \emph{maximum matching} and the \emph{minimum vertex cover}. A matching in a graph is a set of edges without common vertices. A \emph{maximum matching} in a graph is a matching with the largest possible number of edges. A \emph{vertex cover} is a set of vertices such that every edge of the graph is incident to at least one of the set's vertices. A \emph{minimum vertex cover} is a vertex cover with the smallest possible number of vertices. In bipartite graphs, K\H{o}nig’s theorem guarantees that the size of the minimum vertex cover equals the size of the maximum matching \cite{Bondy1976}.

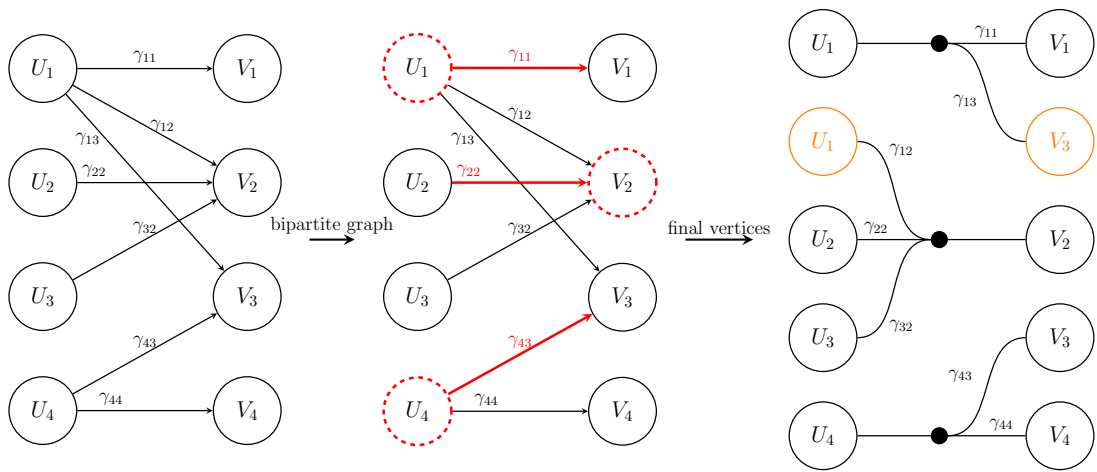
\begin{figure*}[ht]
    \centering
    \begin{tikzpicture}[scale=0.6, transform shape]
        \node[anchor=east] (A) at (0,0){ 
        \begin{tikzpicture}[
    default node/.style={draw, circle, minimum size=1.5cm, font=\Large, thick}, 
    orange node/.style={fill=orange!50}, 
    green node/.style={fill=green!50},  
    label style/.style={font=\normalsize},   
    edge label/.style={font=\large} 
]
\def\vdist{2}
\def\hdist{4}

\node[default node] (U1) {$U_1$};
\node[default node, below=of U1] (U2) {$U_2$};
\node[default node, below=of U2] (U3) {$U_3$};
\node[default node, below=of U3] (U4) {$U_4$};

\node[default node, right= 3 cm of U1] (V1) {$V_1$};
\node[default node, below=of V1] (V2) {$V_2$};
\node[default node, below=of V2] (V3) {$V_3$};
\node[default node, below=of V3] (V4) {$V_4$};

\draw[->] (U1) -- (V1) node[midway, above, edge label] {$\gamma_{11}$};
\draw[->] (U1) -- (V2) node[midway, right, edge label] {$\gamma_{12}$};
\draw[->] (U1) -- (V3) node[near start, left, edge label] {$\gamma_{13}$};
\draw[->] (U2) -- (V2) node[very near start, above, edge label] {$\gamma_{22}$};
\draw[->] (U3) -- (V2) node[midway, above, edge label] {$\gamma_{32}$};
\draw[->] (U4) -- (V3) node[midway, above, edge label] {$\gamma_{43}$};
\draw[->] (U4) -- (V4) node[ near start, above, edge label] {$\gamma_{44}$};

\end{tikzpicture}};
    
        \draw[->, thick] (A.east)++(0.5,0) -- +(1,0) node[midway, above, font=\large] {bipartite graph};
    
        \node[anchor=west] (B) at (2,0) {\begin{tikzpicture}[
    default node/.style={draw, circle, minimum size=1.5cm, font=\Large, thick}, 
    orange node/.style={fill=orange!50}, 
    green node/.style={fill=green!50},  
    label style/.style={font=\normalsize},   
    edge label/.style={font=\large} 
]
\def\vdist{2}
\def\hdist{4}

\node[default node, dashed, draw=red, ultra thick] (U1) {$U_1$};
\node[default node, below=of U1] (U2) {$U_2$};
\node[default node, below=of U2] (U3) {$U_3$};
\node[default node, dashed, draw=red, ultra thick , below=of U3] (U4) {$U_4$};

\node[default node, right= 3 cm of U1] (V1) {$V_1$};
\node[default node, dashed, draw=red, ultra thick, below=of V1] (V2) {$V_2$};
\node[default node, below=of V2] (V3) {$V_3$};
\node[default node, below=of V3] (V4) {$V_4$};

\draw[->, red, ultra thick] (U1) -- (V1) node[midway, above, edge label] {$\gamma_{11}$};
\draw[->] (U1) -- (V2) node[midway, above, edge label] {$\gamma_{12}$};
\draw[->] (U1) -- (V3) node[near start, left, edge label] {$\gamma_{13}$};
\draw[->, red, ultra thick] (U2) -- (V2) node[very near start, above, edge label] {$\gamma_{22}$};
\draw[->] (U3) -- (V2) node[midway, above, edge label] {$\gamma_{32}$};
\draw[->, red, ultra thick] (U4) -- (V3) node[midway, above, edge label] {$\gamma_{43}$};
\draw[->] (U4) -- (V4) node[ near start, above, edge label] {$\gamma_{44}$};

\end{tikzpicture}};

        \draw[->, thick] (B.east)++(0.5,0) -- +(1.5,0) node[midway, above, font=\large] {final vertices};

        \node[anchor=west] (C) at (11,0) {
\begin{tikzpicture}[ 
    default node/.style={draw, circle, minimum size=1.5cm, font=\Large},
    edge label/.style={font=\large},
]
    \def\vdist{4.35cm}
    \def\hdist{5.25}
    \node[draw,circle, default node] (U1) at (0,0*\vdist) {$U_1$};
    \node[draw,circle, default node] (U2) at (0,-1*\vdist) {$U_2$};
    \node[draw,circle, default node] (U3) at (0,-1.5*\vdist) {$U_3$};
    \node[draw,circle, default node] (U4) at (0,-2*\vdist) {$U_4$};
    \node[draw,circle, default node] (V1) at (\hdist,0*\vdist) {$V_1$};
    \node[draw,circle, default node] (V2) at (\hdist,-1*\vdist) {$V_2$};
    \node[draw,circle, default node] (V3) at (\hdist,-1.5*\vdist) {$V_3$};
    \node[draw,circle, default node] (V4) at (\hdist,-2*\vdist) {$V_4$};

    \def\nnodecolor{orange}
    \node[draw,circle,color=\nnodecolor,default node] (U1n) at (0,-0.5*\vdist) {$U_1$};
    
    \node[draw,circle,color=\nnodecolor,default node] (V3n) at (\hdist,-0.5*\vdist) {$V_3$};

    \node[draw,circle,fill=black,inner sep=0, minimum size=10] (b1) at (0.6*\hdist,0*\vdist){};
    \node[draw,circle,fill=black,inner sep=0, minimum size=10] (b2) at (0.6*\hdist,-1*\vdist){};
    \node[draw,circle,fill=black,inner sep=0, minimum size=10] (b3) at (0.6*\hdist,-2*\vdist){};

    \draw (U1) -- (b1) -- (V1) node[midway,above, edge label] {$\gamma_{11}$};
    \draw (b1) to[out=0,in=180] (V3n);
    \node[anchor=south east,  edge label] at (0.82*\hdist,-0.35*\vdist) {$\gamma_{13}$};
    
    \draw (U2) -- (b2) node[xshift=-1.4cm,above,  edge label] {$\gamma_{22}$} -- (V2);
    \draw (U3) to[out=0,in=180] (b2);
    \draw (U1n) to[out=0,in=180] (b2);
    \node[anchor=south east,  edge label] at (0.55*\hdist,-0.6*\vdist) {$\gamma_{12}$};
    \node[anchor=south east,  edge label] at (0.55*\hdist,-1.5*\vdist) {$\gamma_{32}$};
    \draw (U4) -- (b3) -- (V4) node[above,xshift=-1.3cm,  edge label] {$\gamma_{44}$};
    \draw (b3) to[out=0,in=180] (V3);
    \node[anchor=south east,  edge label] at (0.8*\hdist,-1.75*\vdist) {$\gamma_{43}$};
\end{tikzpicture}
};
        
    \end{tikzpicture}
    
    \caption{Illustration of the bipartite graph-based algorithm. The nodes $\{ U_i \}$ and $\{ V_j \}$ represent the non-redundant operators to the left and right of the bond. The edge weights $\gamma_{ij}$ result from the coefficients $\gamma_k$ of the operator strings \eqref{eq:sum_prod_op_repr}. The red dashed vertices form a minimum vertex cover, while the thick red edges form the maximum matching. The orange nodes are required copies since they connect to different vertices.}
    \label{fig:bipartite}
\end{figure*}

We will now briefly review the bipartite-graph-based algorithm of \cite{Ren_2020}. Initially, every term $\hat{O}_k$ of an operator $\hat{O}$ (see Eq.~\eqref{eq:sum_prod_op_repr}) is written as a symbolic operator chain, a state diagram with a chain topology. A key aspect is the vertices connecting each pair of sites. The number of these vertices determines the virtual bond dimension between the final MPO tensors. The algorithm optimizes bonds through a sweeping process from one end of the chain to the other. For each bond, the optimization proceeds in the following steps:
\begin{enumerate}
    \item Create two non-redundant operator chain sets $U$ to the left and $V$ to the right of the to-be-optimized bond. These sets are formed by removing duplicated operator chains within the left and right sides, ignoring the factors $\gamma_k$. The elements of $U$ and $V$ are interpreted as nodes in a bipartite graph.
    \item Create weighted edges between $U$ and $V$ according to the original connectivity. For the $k$-th operator chain with weight $\gamma_k$, assuming that the $i$-th $U$-node matches its left part and the $j$-th $V$ node its right part, an edge with weight $\gamma_{ij} = \gamma_k$ is introduced.
    \item Find a minimum vertex cover $\varepsilon \subset U \cup V$ of the bipartite graph using the Hopcroft–Karp algorithm \cite{Hopcroft1973} and K\H{o}nig's Theorem \cite{Bondy1976}.
    \item Connect the operator chains according to the minimum vertex cover by introducing an operator chain vertex $\nu_\epsilon$ for every $\epsilon \in \varepsilon$.
    \item For every pair $(\epsilon,\epsilon')$ connected by $\nu_\epsilon$ associate $\gamma_{\epsilon\epsilon'}$ with $\epsilon'$. The new factor associated with $\epsilon$ is $1$.
    \item Continue with the next bond.
\end{enumerate}
An example, which was already shown in \cite{Ren_2020}, of the bond optimization is depicted in Fig.~\ref{fig:bipartite}.

\subsubsection{Gamma Interpretation}

We can approach the bond optimization problem from an alternative perspective, which reveals why bipartite graph theory is insufficient for determining optimal bond dimensions in the case of related coefficient $\gamma_k$. Again, we create the non-redundant operator chain sets $U$ and $V$. Now, we represent the connection between two sites as a matrix
\begin{equation}\label{eq:gamma_matrix}
    \Gamma = 
    \begin{array}{c|cccccc|}
    & \text{V}_1 & \text{V}_2 & \ldots & \text{V}_j & \ldots & \text{V}_m \\
    \hline
    \text{U}_1 & \gamma_{11} & \gamma_{12} & \ldots & \gamma_{1j} & \ldots & \gamma_{1m} \\
    \text{U}_2 & \gamma_{21} & \gamma_{22} & \ldots & \gamma_{2j} & \ldots & \gamma_{2m} \\
    \vdots & \vdots & \vdots &  & \vdots &  & \vdots \\
    \text{U}_i & \gamma_{i1} & \gamma_{i2} & \ldots & \gamma_{ij} & \ldots & \gamma_{im} \\
    \vdots & \vdots & \vdots &  & \vdots &  & \vdots \\
    \text{U}_n & \gamma_{n1} & \gamma_{n2} & \ldots & \gamma_{nj} & \ldots & \gamma_{nm} \\
    \hline
\end{array}\, ,
\end{equation}\\
where rows and columns corresponds to the nodes in \( U \) and \( V \) respectively. The \( \Gamma \) matrix can be seen as a modified adjacency matrix, reflecting the connectivity of the bipartite graph. An entry \( \gamma_{ij} \) in the matrix indicates the connectivity between the \( i \)-th \( U \)-node and the \( j \)-th \( V \)-node. The values of these entries correspond to the \( \gamma \)-coefficients associated with the given edge.

The minimum possible dimension for the bond represented by $\Gamma$ in Eq.~\eqref{eq:gamma_matrix} is $\rank(\Gamma)$. Thus the problem of bond dimension optimization reduces to finding $\{ \alpha_{ri} \}$ and $\{ \beta_{rj} \}$ such that
\begin{equation}
\setlength{\arraycolsep}{2pt}
\Gamma = \sum_{r=1}^R
\begin{pmatrix}
    \alpha_{r1} \\
    \alpha_{r2} \\
    \vdots \\
    \alpha_{rn} 
\end{pmatrix}
\begin{pmatrix}
    \beta_{r1} & \beta_{r2} & \ldots & \beta_{rm}
\end{pmatrix} 
\setlength{\arraycolsep}{5pt}
\end{equation}
and $R$ is as close to $\rank(\Gamma)$ as possible. One could use a singular value decomposition (SVD) for a set of known values of the factors $\gamma_{ij}$. However, we want to find a symbolic representation for arbitrary values. Another problem of the SVD is a possible failure of convergence.

The bipartite graph algorithm simplifies the problem by imposing additional constraints on the terms. More specifically, one of the vectors (either row or column) in every term must be a unit vector with a single non-zero element. In other words, the algorithm selects specific rows or columns directly from the \( \Gamma \) matrix. To elaborate on this operation in more detail, let us examine the following matrix equivalent to the example shown in Fig.~\ref{fig:bipartite}:
\begin{equation}
    \Gamma_e = 
\begin{pmatrix}
    \gamma_{11} & \gamma_{12} & \gamma_{13} & 0 \\
    0      & \gamma_{22} & 0      & 0 \\
    0      & \gamma_{32} & 0      & 0 \\
    0      & 0      & \gamma_{43} & \gamma_{44}
\end{pmatrix}
\end{equation}
For this toy example, selecting rows $1$ and $4$ as well as column $2$ via the unit vectors suffices to decompose the matrix as
\begin{equation}
\begin{split}
    \Gamma_e &= 
    \begin{pmatrix}
        \textcolor{red}{\gamma_{11}} & \textcolor{violet}{\gamma_{12}} & \textcolor{red}{\gamma_{13}} & \textcolor{red}0 \\
        0      & \textcolor{violet}{\gamma_{22}} & 0      & 0 \\
        0      & \textcolor{violet}{\gamma_{32}} & 0      & 0 \\
        \textcolor{blue}0      & \textcolor{violet}0      & \textcolor{blue}{\gamma_{43}} & \textcolor{blue}{\gamma_{44}}
    \end{pmatrix} \\
    &= \begin{pmatrix}
        \gamma_{12} \\
        \gamma_{22} \\
        \gamma_{32} \\
        0
    \end{pmatrix}
    \begin{pmatrix}
        0 & \textcolor{violet}1 & 0 & 0
    \end{pmatrix} 
    +
    \begin{pmatrix}
        \textcolor{red}1 \\
        0 \\
        0 \\
        0 
    \end{pmatrix}
    \begin{pmatrix}
        \gamma_{11} & 0 & \gamma_{13} & 0
    \end{pmatrix}\\
    & \qquad +
    \begin{pmatrix}
        0 \\
        0 \\
        0 \\
        \textcolor{blue}1 
    \end{pmatrix}
    \begin{pmatrix}
        0 & 0 & \gamma_{43} & \gamma_{44}
    \end{pmatrix} \, .
\end{split}
\end{equation}

As observed, we arrive at a rank-$3$ solution simply by selecting specific rows and columns. This is equivalent to the solution found via the minimum vertex cover in Fig.~\ref{fig:bipartite}. This method turns out to work well in many cases. However, we can now see that it must be sub-optimal if a simultaneous combination of rows and columns is required. A minimal example where the non-optimality becomes evident is given in Appendix~\ref{app:minimum_example_bp_not_optimal}, and we provide a more complicated example in Appendix~\ref{app:SGE}. This limitation becomes particularly problematic in scenarios with uniform coefficients, such as those found in lattice models in physics. Such patterns, while uncommon in usual chemical models \cite{10.1063/5.0218773}, are crucial when many Hamiltonian terms share the same coefficient. Ignoring these patterns can lead to sub-optimal representations, as shown in Sec.~\ref{sec:results} and most notably in Sec.~\ref{sec:appl_molecules}. This underscores the need for an enhanced algorithm that addresses these cases effectively.
\begin{figure*}
    \centering
\begin{tikzpicture}[scale=0.6, transform shape]

    \node (A) {\begin{tikzpicture}[
    default node/.style={draw, circle, minimum size=1.5cm, font=\Large, thick}, 
    orange node/.style={fill=orange!50}, 
    green node/.style={fill=green!50},  
    edge label/.style={font=\Large}, 
    node distance = 1.5
]
\def\vdist{2}
\def\hdist{4}

\node[default node] (U1) {$U_1$};
\node[default node, below=of U1] (U2) {$U_2$};
\node[default node, below=of U2] (U3) {$U_3$};
\node[default node, below=of U3] (U4) {$U_4$};
\node[default node, below=of U4] (U5) {$U_5$};

\node[default node, right= 4 cm of U1] (V1) {$V_1$};
\node[default node, below=of V1] (V2) {$V_2$};
\node[default node, below=of V2] (V3) {$V_3$};
\node[default node, below=of V3] (V4) {$V_4$};
\node[default node, below=of V4] (V5) {$V_5$};

\draw[->] (U1) -- (V1) node[midway, above, edge label] {$3\gamma_{11}$};
\draw[->] (U1) -- (V2) node[midway, right, edge label] {$3\gamma_{12}$};
\draw[->] (U1) -- (V3) node[near start, right, edge label] {$-\gamma_{13}$};
\draw[->] (U1) -- (V5) node[pos = 0.08, left, edge label] {$2\gamma_{12}$};
\draw[->] (U2) -- (V2) node[ near start, above, edge label] {$3\gamma_{22}$};
\draw[->] (U2) -- (V5) node[very near start, left, edge label] {$2\gamma_{22}$};
\draw[->] (U3) -- (V2) node[midway, above, edge label] {$3\gamma_{32}$};
\draw[->] (U3) -- (V5) node[very near start, left, edge label] {$2\gamma_{32}$};

\draw[->] (U4) -- (V3) node[very near start, above, edge label] {$2\gamma_{43}$};
\draw[->] (U4) -- (V4) node[ near start, above, edge label] {$2\gamma_{44}$};
\draw[->] (U5) -- (V3) node[near start, left, edge label] {$-\gamma_{43}$};
\draw[->] (U5) -- (V4) node[midway, above, edge label] {$-\gamma_{44}$};

\end{tikzpicture}
};
    \node[right= 2 cm of A] (B) {\begin{tikzpicture}[
    default node/.style={draw, circle, minimum size=1.5cm, font=\Large}, 
    orange node/.style={fill=orange!50}, 
    green node/.style={fill=green!50},  
    label style/.style={font=\small},   
    edge label/.style={font=\Large}, 
    node distance = 1.5cm
]
\def\vdist{2}
\def\hdist{4}

\node[default node] (U1) {$U_1$};
\node[default node, below=of U1] (U2) {$U_2$};
\node[default node, below=of U2] (U3) {$U_3$};
\node[default node, below=of U3] (U4) {$U_4$};
\node[default node, below=of U4] (U5) {$U_5$};

\node[default node, orange node, right=of U1] (U1T) {$\tilde{U}_1$};
\node[default node, orange node, right=of U2] (U2T) {$\tilde{U}_2$};
\node[default node, orange node, right=of U3] (U3T) {$\tilde{U}_3$};
\node[default node, orange node, right=of U4] (U4T) {$\tilde{U}_4$};

\node[default node, green node, right=of U1T] (V1T) {$\tilde{V}_1$};
\node[default node, green node, right=of U2T] (V2T) {$\tilde{V}_2$};
\node[default node, green node, right=of U3T] (V3T) {$\tilde{V}_3$};
\node[default node, green node, right=of U4T] (V4T) {$\tilde{V}_4$};

\node[default node, right=of V1T] (V1) {$V_1$};
\node[default node, below=of V1] (V2) {$V_2$};
\node[default node, below=of V2] (V3) {$V_3$};
\node[default node, below=of V3] (V4) {$V_4$};
\node[default node, below=of V4] (V5) {$V_5$};

\foreach \i in {1,2,3,4} {
    \draw[->] (U\i) -- (U\i T) node[midway, above, edge label] {1};
}
\draw[->] (U5) -- (U4T) node[near start,yshift=0.4cm, above, edge label] {$-\frac{1}{2}$};

\draw[->] (U1T) -- (V1T) node[midway, above, edge label] {$3\gamma_{11}$};
\draw[->] (U1T) -- (V2T) node[midway, right, edge label] {$3\gamma_{12}$};
\draw[->] (U1T) -- (V3T) node[near start, left, edge label] {$-\gamma_{13}$};
\draw[->] (U2T) -- (V2T) node[near end, above, edge label] {3$\gamma_{22}$};
\draw[->] (U3T) -- (V2T) node[midway, left, edge label] {3$\gamma_{32}$};
\draw[->] (U4T) -- (V3T) node[midway, left, edge label] {2$\gamma_{43}$};
\draw[->] (U4T) -- (V4T) node[midway, above, edge label] {2$\gamma_{44}$};

\foreach \i in {1,2,3,4} {
    \draw[->] (V\i T) -- (V\i) node[midway, above, edge label] {1};
}
\draw[->] (V2T) -- (V5) node[midway, right, edge label] {$\frac{2}{3}$};

\end{tikzpicture}
};
    \node[below =2 cm of B] (C) {\begin{tikzpicture}[
    default node/.style={draw, circle, minimum size=1.5cm, font=\Large}, 
    orange node/.style={fill=orange!50}, 
    green node/.style={fill=green!50},  
    label style/.style={font=\small},   
    edge label/.style={font=\Large}, 
    node distance=1.5cm
]
\def\vdist{3}
\def\hdist{4}

\node[default node] (U1) {$U_1$};
\node[default node, below=of U1] (U2) {$U_2$};
\node[default node, below=of U2] (U3) {$U_3$};
\node[default node, below=of U3] (U4) {$U_4$};
\node[default node, below=of U4] (U5) {$U_5$};

\node[default node, dashed, draw=red, orange node, ultra thick, right=of U1] (U1T) {$\tilde{U}_1$};
\node[default node, orange node, right=of U2] (U2T) {$\tilde{U}_2$};
\node[default node, orange node, right=of U3] (U3T) {$\tilde{U}_3$};
\node[default node, dashed, draw=red, ultra thick, orange node, right=of U4] (U4T) {$\tilde{U}_4$};

\node[default node, green node, right=of U1T] (V1T) {$\tilde{V}_1$};
\node[default node, dashed, draw=red, ultra thick, green node,  right=of U2T] (V2T) {$\tilde{V}_2$};
\node[default node, green node, right=of U3T] (V3T) {$\tilde{V}_3$};
\node[default node, green node, right=of U4T] (V4T) {$\tilde{V}_4$};

\node[default node, right=of V1T] (V1) {$V_1$};
\node[default node, below=of V1] (V2) {$V_2$};
\node[default node, below=of V2] (V3) {$V_3$};
\node[default node, below=of V3] (V4) {$V_4$};
\node[default node, below=of V4] (V5) {$V_5$};

\foreach \i in {1,2,3,4} {
    \draw[->] (U\i) -- (U\i T) node[midway, above, edge label] {1};
}
\draw[->] (U5) -- (U4T) node[near start,xshift=-0.4cm, above, edge label] {$-\frac{1}{2}$};
\draw[->, red, ultra thick] (U1T) -- (V1T) node[midway, above, edge label] {$3\gamma_{11}$};
\draw[->] (U1T) -- (V2T) node[midway, right, edge label] {$3\gamma_{12}$};
\draw[->] (U1T) -- (V3T) node[near start, left, edge label] {$-\gamma_{13}$};
\draw[->, red, ultra thick] (U2T) -- (V2T) node[near end, above, edge label] {$3\gamma_{22}$};
\draw[->] (U3T) -- (V2T) node[midway, left, edge label] {$3\gamma_{32}$};
\draw[->, red, ultra thick] (U4T) -- (V3T) node[midway, left, edge label] {$2\gamma_{43}$};
\draw[->] (U4T) -- (V4T) node[midway, above, edge label] {$2\gamma_{44}$};

\foreach \i in {1,2,3,4} {
    \draw[->] (V\i T) -- (V\i) node[midway, above, edge label] {1};
}
\draw[->] (V2T) -- (V5) node[midway, right, edge label] {$\frac{2}{3}$};

\end{tikzpicture}};
    \node[left= 2 cm of C] (D) {\begin{tikzpicture}[ 
    default node/.style={draw, circle, minimum size=1.5cm, font=\Large},
    edge label/.style={font=\Large},
]
    \def\vdist{4.83cm}
    \def\hdist{5.25}
    \node[draw,circle, default node] (U1) at (0,0*\vdist) {$U_1$};
    \node[draw,circle, default node] (U2) at (0,-1*\vdist) {$U_2$};
    \node[draw,circle, default node] (U3) at (0,-1.5*\vdist) {$U_3$};
    \node[draw,circle, default node] (U4) at (0,-2*\vdist) {$U_4$};
    \node[draw,circle, default node] (U5) at (0,-2.5*\vdist) {$U_5$};
    \node[draw,circle, default node] (V1) at (\hdist,0*\vdist) {$V_1$};
    \node[draw,circle, default node] (V2) at (\hdist,-1*\vdist) {$V_2$};
    \node[draw,circle, default node] (V3) at (\hdist,-2*\vdist) {$V_3$};
    \node[draw,circle, default node] (V4) at (\hdist,-2.5*\vdist) {$V_4$};
    \node[draw,circle, default node] (V5) at (\hdist,-1.5*\vdist) {$V_5$};

    \def\nnodecolor{orange}
    \node[draw,circle,color=\nnodecolor,default node] (U1n) at (0,-0.5*\vdist) {$U_1$};
    
    \node[draw,circle,color=\nnodecolor,default node] (V3n) at (\hdist,-0.5*\vdist) {$V_3$};

    \node[draw,circle,fill=black,inner sep=0, minimum size=10] (b1) at (0.5*\hdist,0*\vdist){};
    \node[draw,circle,fill=black,inner sep=0, minimum size=10] (b2) at (0.5*\hdist,-1*\vdist){};
    \node[draw,circle,fill=black,inner sep=0, minimum size=10] (b3) at (0.5*\hdist,-2*\vdist){};

    \draw (U1) -- (b1) -- (V1) node[midway,above, edge label] {$3\gamma_{11}$};
    \draw (b1) to[out=0,in=180] (V3n);
    \node[anchor=south east,  edge label] at (0.82*\hdist,-0.35*\vdist) {-$\gamma_{13}$};
    
    \draw (U2) -- (b2) node[xshift=-2cm,above,  edge label] {3$\gamma_{22}$} -- (V2);
    \draw (U3) to[out=0,in=180] (b2);
    \draw (U1n) to[out=0,in=180] (b2);
    \node[anchor=south east,  edge label] at (0.3*\hdist,-0.6*\vdist) {$3\gamma_{12}$};
    \draw (b2) to[out=0,in=180] (V5);
    \node[anchor=south east,  edge label] at (0.4*\hdist,-1.5*\vdist) {$3\gamma_{32}$};
    \node[anchor=south east,  edge label] at (0.75*\hdist,-1.42*\vdist) {$\frac{2}{3}$};
    \draw (U4) -- (b3) -- (V3) node[above,xshift=-1.5cm,  edge label] {$2\gamma_{43}$};
    \draw (U5) to[out=0,in=180] (b3);
    \draw (b3) to[out=0,in=180] (V4);
    \node[anchor=south east,  edge label] at (0.15*\hdist,-2.4*\vdist) {$-\frac{1}{2}$};
    \node[anchor=south east,  edge label] at (0.82*\hdist,-2.33*\vdist) {2$\gamma_{44}$};
\end{tikzpicture}
};
    
    \draw[->, very thick] (A.east) -- (B.west) node[midway, above, font=\large] {step 1}; 
    \draw[->, very thick, shorten >=10pt] (B.south) -- (C.north) node[midway, right, font=\large] {step 2}; 
    \draw[->, very thick] (C.west) -- (D.east) node[midway, below, font=\large] {step 3}; 
    
\end{tikzpicture}
\caption{The figure illustrates our proposed algorithm. In step 1, the SGE is applied to create virtual layers. The orange nodes represent the virtual nodes $\tilde{U}$ on the left-hand side, while the green nodes represent the virtual nodes $\tilde{V}$ on the right-hand side. Step 2 finds the minimum vertex covering of the inner bipartite graph given by the red dashed vertices. The thick red edges correspond to the maximum matching. Finally, step 3 leads to the final connectivity and factor association.}
\label{fig:algorithm}
\end{figure*}
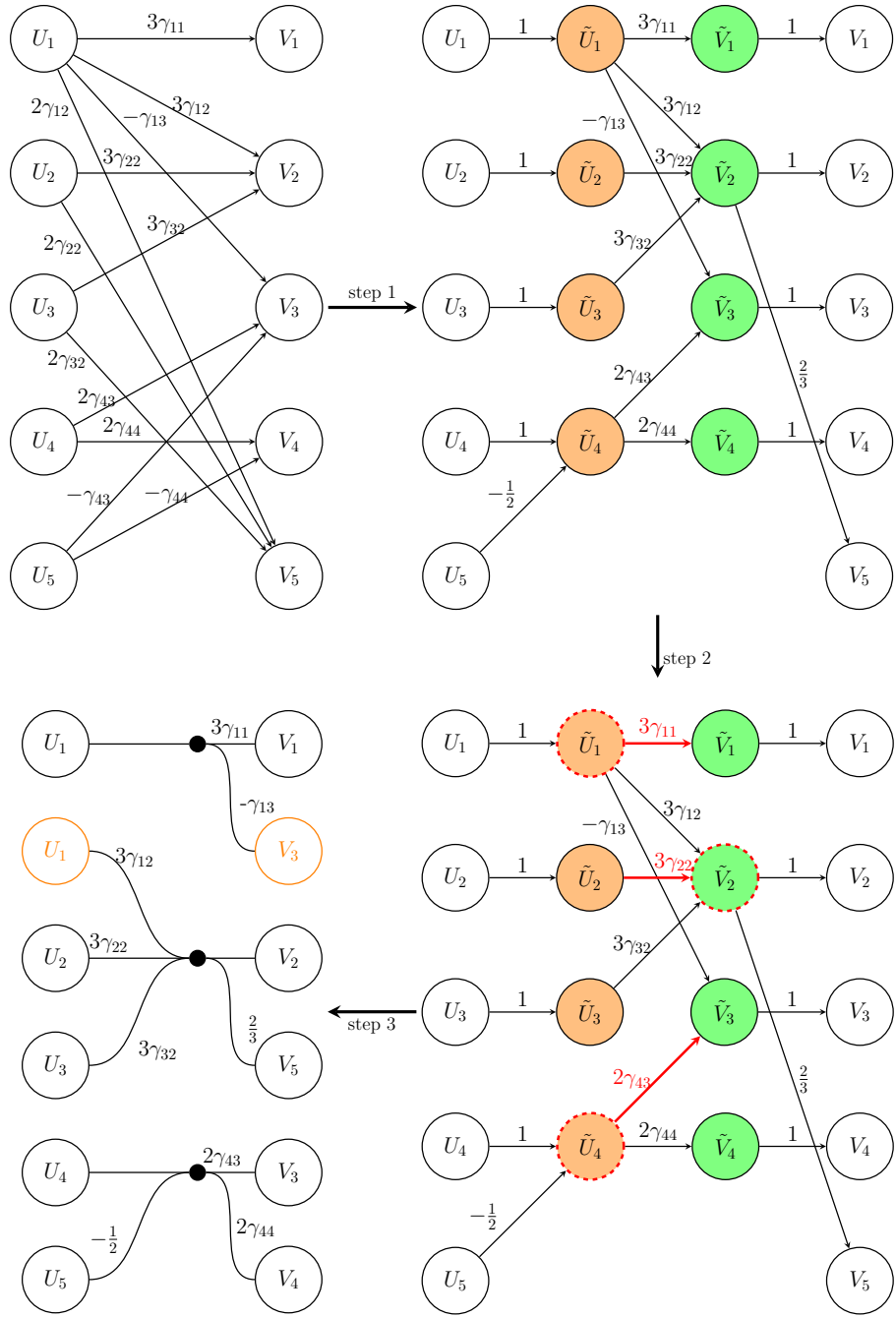

\subsubsection{Proposed Improvement: Symbolic Gaussian Elimination}\label{sec:SGE}

We propose a pre-processing step to the bipartite-graph-based algorithm to overcome its limitations. This pre-processing step is a variant of the Gaussian elimination procedure applied to the \( \Gamma \)-matrix to eliminate linearly dependent rows or columns while avoiding complicated symbolic expressions.

\begin{algorithm}[ht]
\RestyleAlgo{ruled}
\caption{Symbolic Gaussian Elimination and Bipartite Graph Optimization}
\label{alg:algorithm}
\SetKwInOut{Input}{Input}
\SetKwInOut{Output}{Output}

\Input{Matrix $\Gamma$ with coefficients $\gamma_{ij}$}
\Output{Updated matrix $\tilde{\Gamma}$ }

\textbf{Initialize} two identity matrices $A$ and $B$ matching the row \& column dimensions of $\Gamma$\;

\emph{Detect} parallel rows and columns of $\Gamma$ and \emph{De-parallelize} them.
    

\For{$t \leftarrow 1$ \KwTo maxIter} {
\ForEach{row $i$ in $\Gamma$}{
    \ForEach{column $j$ in $\Gamma$}{
        \If{$\gamma_{ij}$ is non-zero}{
            Apply \textit{restricted} row elimination on row $i$\;
            Update $A$ to track row operations\;
            Apply \textit{restricted} column elimination on column $j$\;
            Update $B$ to track column operations\;
        }
    }
    
}
\ForEach{row $i$ in $\Gamma$}{
    \If{row $i$ in $\Gamma$ is zero}{
        Remove row $i$ from $\Gamma$ and the corresponding column $i$ from $B$\;
    }
    }
    \ForEach{column $j$ in $\Gamma$}{
    \If{column $j$ in $\Gamma$ is zero}{
        Remove column $j$ from $\Gamma$ and the corresponding row $j$ from $A$\;
    }
    }

\If{convergence criteria met (e.g., no changes in $\tilde{\Gamma}$ or maximum optimization $maxIter$ reached)}{
        \textbf{Break} the loop\;
}
}

Apply bipartite graph algorithm to the reduced $\tilde{\Gamma}$ to determine minimum vertex cover\;
\If{minimum vertex cover is bigger than $\Gamma$}{
   Replace $\Tilde{\Gamma}$ with $\Gamma$;
}

\Output{ $\tilde{\Gamma}$, $A$, and $B$ matrices}
\end{algorithm}

The proposed symbolic adaptation of Gaussian elimination is referred to as \emph{Symbolic Gaussian Elimination} (SGE). In this method, the entries of the matrix $\Gamma$ are stored symbolically as $(c, s_i)$, where $c \in \Q$ and $s_i$ is a symbol from a set of symbols $\mathcal{S}$ that can be mapped to a numeric value at a later point. Accordingly, the row and column elimination steps are also symbolic, and we must enforce some limitations to keep the method efficient.

Firstly, we do not allow additional symbols to be introduced into $\mathcal{S}$ during the elimination. Notably, this disallows sums of different symbolic values, i.e.,
\begin{equation}
    (c_1, s_i) + (c_2, s_j) = \begin{cases}
        (c_1 + c_2, s_i) & \text{if} \quad i = j \\
        \text{prohibited} & \text{if} \quad i \neq j.
    \end{cases}
\end{equation}
This restriction simplifies the implementation for subsequent iterations, ensuring that the symbolic nature of the process is preserved and keeping the method practical and easily implementable. The resulting row and column operations are called \emph{restricted row and column operations}, as they follow the standard principles of Gaussian elimination but with additional constraints.

The second limitation is demanding $c \in \Q$. This constraint ensures mathematical stability while offering sufficient representation power. We noticed that limiting the coefficients to integers would severely restrict the representation capabilities of the matrix, thus leaving some simple cases unsolvable by the algorithm. By allowing fractional real numbers, we significantly increase the representation scope while maintaining stable calculations. Although this approach imposes some limitations, it provides a practical balance, covering most cases effectively and ensuring the process remains stable.

To keep track of the operations performed on $\Gamma$, we introduce the two matrices \( A \) and \( B \) and initialize them as the identity matrix. They are updated in tandem with $\Gamma$. In the end
each column $j$ in \( A \) and each row $i$ in \( B \) represents the inverse of the operations applied to the \( j \)-th row of \( \Gamma \) and the \( i \)-th column of the \( \Gamma \) respectively. This ensures
\begin{equation}
\Gamma = A \cdot \Tilde{\Gamma} \cdot B
\end{equation}
holds, where \( \Tilde{\Gamma} \) is the updated \( \Gamma \)-matrix. If a zero row or zero column is encountered in the \( \Tilde{\Gamma} \)-matrix, we eliminate it and the corresponding column from \( A \) or the corresponding row from \( B \), respectively. Eventually the SGE yields \( \Tilde{\Gamma} \).

As a second step, we use the bipartite graph method to minimize the number of vertices that correspond to the updated matrix $\Tilde{\Gamma}$. Note that this additional step is redundant if the entries in the \( \Gamma \)-matrix have uniform symbolic coefficients $\gamma_{ij}$. In cases of partial uniformity among the coefficients, the bipartite graph algorithm remains essential for detecting the minimal rank.
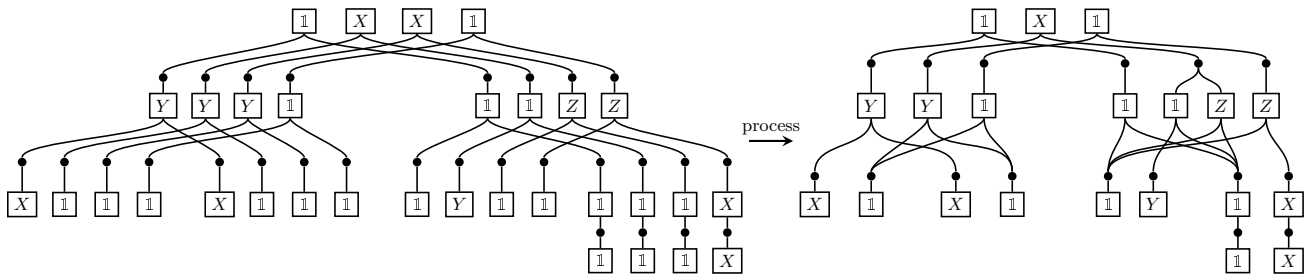
\begin{figure*}[ht]
    \centering
    
    \begin{tikzpicture}[scale=0.75, transform shape]
    \node (A) at (0,0) {\begin{tikzpicture}[baseline=(current bounding box.center), scale=1]
\node[draw] (HE1) at (-0.5,0){$\mathbbm{1}$};
\node[draw] (HE2) at (0.5,0){$X$};
\node[draw] (HE3) at (1.5,0){$X$};
\node[draw] (HE4) at (2.5,0){$\mathbbm{1}$};

\node[fill, circle, scale=0.5] (V51) at (2.75,-1){};		
\node[fill, circle, scale=0.5] (V52) at (3.5,-1){};
\node[fill, circle, scale=0.5] (V53) at (4.25,-1){};		
\node[fill, circle, scale=0.5] (V54) at (5,-1){};		

\node[fill, circle, scale=0.5] (V21) at (-3,-1){};
\node[fill, circle, scale=0.5] (V22) at (-2.25,-1){};
\node[fill, circle, scale=0.5] (V23) at (-1.5,-1){};
\node[fill, circle, scale=0.5] (V24) at (-0.75,-1){};

\node[draw] (HE5) at (-3,-1.5){$Y$};
\node[draw] (HE6) at (-2.25,-1.5){$Y$};
\node[draw] (HE7) at (-1.5,-1.5){$Y$};
\node[draw] (HE8) at (-0.75,-1.5){$\mathbbm{1}$};

\node[draw] (HE9) at (2.75,-1.5){$\mathbbm{1}$};
\node[draw] (HE10) at (3.5,-1.5){$\mathbbm{1}$};
\node[draw] (HE11) at (4.25,-1.5){$Z$};
\node[draw] (HE12) at (5,-1.5){$Z$};

\node[fill, circle, scale=0.5] (V31) at (-5.5,-2.5){};
\node[fill, circle, scale=0.5] (V32) at (-4.75,-2.5){};
\node[fill, circle, scale=0.5] (V33) at (-4,-2.5){};
\node[fill, circle, scale=0.5] (V34) at (-3.25,-2.5){};

\node[fill, circle, scale=0.5] (V41) at (-2,-2.5){};
\node[fill, circle, scale=0.5] (V42) at (-1.25,-2.5){};
\node[fill, circle, scale=0.5] (V43) at (-0.5,-2.5){};
\node[fill, circle, scale=0.5] (V44) at (0.25,-2.5){};

\node[fill, circle, scale=0.5] (V61) at (1.5,-2.5){};
\node[fill, circle, scale=0.5] (V62) at (2.25,-2.5){};
\node[fill, circle, scale=0.5] (V63) at (3,-2.5){};
\node[fill, circle, scale=0.5] (V64) at (3.75,-2.5){};

\node[fill, circle, scale=0.5] (V71) at (4.75,-2.5){};
\node[fill, circle, scale=0.5] (V72) at (5.5,-2.5){};
\node[fill, circle, scale=0.5] (V73) at (6.25,-2.5){};
\node[fill, circle, scale=0.5] (V74) at (7,-2.5){};

\node[draw] (HE13) at (-5.5,-3.25){$X$};
\node[draw] (HE14) at (-4.75,-3.25){$\mathbbm{1}$};
\node[draw] (HE15) at (-4,-3.25){$\mathbbm{1}$};
\node[draw] (HE16) at (-3.25,-3.25){$\mathbbm{1}$};

\node[draw] (HE17) at (-2,-3.25){$X$};
\node[draw] (HE18) at (-1.25,-3.25){$\mathbbm{1}$};
\node[draw] (HE19) at (-0.5,-3.25){$\mathbbm{1}$};
\node[draw] (HE20) at (0.25,-3.25){$\mathbbm{1}$};

\node[draw] (HE21) at (1.5,-3.25){$\mathbbm{1}$};
\node[draw] (HE22) at (2.25,-3.25){$Y$};
\node[draw] (HE23) at (3,-3.25){$\mathbbm{1}$};
\node[draw] (HE24) at (3.75,-3.25){$\mathbbm{1}$};

\node[draw] (HE25) at (4.75,-3.25){$\mathbbm{1}$};
\node[draw] (HE26) at (5.5,-3.25){$\mathbbm{1}$};
\node[draw] (HE27) at (6.25,-3.25){$\mathbbm{1}$};
\node[draw] (HE28) at (7,-3.25){$X$};

\node[fill, circle, scale=0.5] (V81) at (4.75,-3.75){};
\node[fill, circle, scale=0.5] (V82) at (5.5,-3.75){};
\node[fill, circle, scale=0.5] (V83) at (6.25,-3.75){};
\node[fill, circle, scale=0.5] (V84) at (7,-3.75){};

\node[draw] (HE29) at (4.75,-4.25){$\mathbbm{1}$};
\node[draw] (HE30) at (5.5,-4.25){$\mathbbm{1}$};
\node[draw] (HE31) at (6.25,-4.25){$\mathbbm{1}$};
\node[draw] (HE32) at (7,-4.25){$X$};

\draw (HE1) to [out=-90, in=90, looseness=0.2] (V21) -- (HE5);
\draw (HE2) to [out=-90, in=90, looseness=0.2] (V22) -- (HE6);
\draw (HE3) to [out=-90, in=90, looseness=0.2] (V23) -- (HE7);
\draw (HE4) to [out=-90, in=90, looseness=0.2] (V24) -- (HE8);

\draw (HE1) to [out=-90, in=90, looseness=0.2] (V51) -- (HE9);
\draw (HE2) to [out=-90, in=90, looseness=0.2] (V52) -- (HE10);
\draw (HE3) to [out=-90, in=90, looseness=0.2] (V53) -- (HE11);
\draw (HE4) to [out=-90, in=90, looseness=0.2] (V54) -- (HE12);

\draw (HE5) to [out=-90, in=90, looseness=0.3] (V31) -- (HE13);
\draw (HE6) to [out=-90, in=90, looseness=0.3] (V32) -- (HE14);
\draw (HE7) to [out=-90, in=90, looseness=0.3] (V33) -- (HE15);
\draw (HE8) to [out=-90, in=90, looseness=0.3] (V34) -- (HE16);

\draw (HE5) to [out=-90, in=90, looseness=0.3] (V41) -- (HE17);
\draw (HE6) to [out=-90, in=90, looseness=0.3] (V42) -- (HE18);
\draw (HE7) to [out=-90, in=90, looseness=0.3] (V43) -- (HE19);
\draw (HE8) to [out=-90, in=90, looseness=0.3] (V44) -- (HE20);

\draw (HE9) to [out=-90, in=90, looseness=0.3] (V61) -- (HE21);
\draw (HE10) to [out=-90, in=90, looseness=0.3] (V62) -- (HE22);
\draw (HE11) to [out=-90, in=90, looseness=0.3] (V63) -- (HE23);
\draw (HE12) to [out=-90, in=90, looseness=0.3] (V64) -- (HE24);

\draw (HE9) to [out=-90, in=90, looseness=0.3] (V71) -- (HE25);
\draw (HE10) to [out=-90, in=90, looseness=0.3] (V72) -- (HE26);
\draw (HE11) to [out=-90, in=90, looseness=0.3] (V73) -- (HE27);
\draw (HE12) to [out=-90, in=90, looseness=0.3] (V74) -- (HE28);

\draw (HE25) to [out=-90, in=90, looseness=0.2] (V81) -- (HE29);
\draw (HE26) to [out=-90, in=90, looseness=0.2] (V82) -- (HE30);
\draw (HE27) to [out=-90, in=90, looseness=0.2] (V83) -- (HE31);
\draw (HE28) to [out=-90, in=90, looseness=0.2] (V84) -- (HE32);

\end{tikzpicture}
};
    \node (B) at(12,0) {\begin{tikzpicture}
\node[draw] (HE1) at (0,0){$\mathbbm{1}$};
\node[fill, circle, scale=0.5] (V11) at (2.5,-0.75){};		
\node[fill, circle, scale=0.5] (V21) at (-2,-0.75){};

\node[draw] (HE2) at (1,0){$X$};
\node[fill, circle, scale=0.5] (V12) at (3.8,-0.75){};
\node[fill, circle, scale=0.5] (V22) at (-1,-0.75){};

\node[draw] (HE3) at (2,0){$\mathbbm{1}$};
\node[fill, circle, scale=0.5] (V13) at (5,-0.75){};
\node[fill, circle, scale=0.5] (V23) at (0,-0.75){};

\node[draw] (HE4) at (-2,-1.5){$Y$};
\node[fill, circle, scale=0.5] (V41) at (-3,-2.75){};
\node[fill, circle, scale=0.5] (V51) at (-0.5,-2.75){};

\node[draw] (HE5) at (-1,-1.5){$Y$};
\node[fill, circle, scale=0.5] (V42) at (-2,-2.75){};

\node[draw] (HE6) at (0,-1.5){$\mathbbm{1}$};
\node[fill, circle, scale=0.5] (V52) at (0.5,-2.75){};

\node[draw] (L1) at (-3,-3.25){$X$};
\node[draw] (L2) at (-2,-3.25){$\mathbbm{1}$};

\node[draw] (L3) at (-0.5,-3.25){$X$};
\node[draw] (L4) at (0.5,-3.25){$\mathbbm{1}$};

\node[draw] (HE7) at (2.5,-1.5){$\mathbbm{1}$};
\node[draw] (HE8) at (3.4,-1.5){$\mathbbm{1}$};
\node[draw] (HE9) at (4.2,-1.5){$Z$};
\node[draw] (HE10) at (5,-1.5){$Z$};

\node[fill, circle, scale=0.5] (V61) at (2.2,-2.75){};
\node[fill, circle, scale=0.5] (V62) at (3,-2.75){};

\node[draw] (HE11) at (2.2,-3.25){$\mathbbm{1}$};
\node[draw] (HE12) at (3,-3.25){$Y$};

\node[fill, circle, scale=0.5] (V71) at (4.5,-2.75){};
\node[fill, circle, scale=0.5] (V72) at (5.4,-2.75){};

\node[draw] (HE13) at (4.5,-3.25){$\mathbbm{1}$};
\node[draw] (HE14) at (5.4,-3.25){$X$};

\node[fill, circle, scale=0.5] (V81) at (4.5,-3.75){};
\node[fill, circle, scale=0.5] (V82) at (5.4,-3.75){};

\node[draw] (L5) at (4.5,-4.25){$\mathbbm{1}$};
\node[draw] (L6) at (5.4,-4.25){$X$};

\draw (HE1) to [out=-90, in=90, looseness=0.3] (V21) -- (HE4);
\draw (HE2) to [out=-90, in=90, looseness=0.3] (V22) -- (HE5);
\draw (HE3) to [out=-90, in=90, looseness=0.3] (V23) -- (HE6);

\draw (HE1) to [out=-90, in=90, looseness=0.2] (V11) -- (HE7);
\draw (HE2) to [out=-90, in=90, looseness=0.2] (V12) to [out=-90, in=90] (HE8);
\draw (V12) to [out=-90, in=90] (HE9);
\draw (HE3) to [out=-90, in=90, looseness=0.2] (V13) -- (HE10);

\draw (HE4) to [out=-90, in=90] (V41) -- (L1);
\draw (HE5) to [out=-90, in=90, looseness=0.4] (V42) -- (L2);
\draw (HE6) to [out=-90, in=90, looseness=0.3] (V42);

\draw (HE4) to [out=-90, in=90] (V51) -- (L3);
\draw (HE5) to [out=-90, in=90] (V52) -- (L4);
\draw (HE6) to [out=-90, in=90] (V52);

\draw (HE7) to [out=-90, in=90] (V61) -- (HE11);
\draw (HE8) to [out=-90, in=90] (V62) -- (HE12);
\draw (HE9) to [out=-90, in=90, looseness=0.7] (V61);
\draw (HE10) to [out=-90, in=90, looseness=0.5] (V61);

\draw (HE7) to [out=-90, in=90, looseness=0.5] (V71) -- (HE13);
\draw (HE8) to [out=-90, in=90] (V71);
\draw (HE9) to [out=-90, in=90] (V71);
\draw (HE10) to [out=-90, in=90]  (V72) -- (HE14);

\draw (HE13) -- (L5);
\draw (HE14) -- (L6);
\end{tikzpicture}
};
    
    \draw[->, thick] (A.east) -- (B.west) node[midway, above] {process}; 
    
\end{tikzpicture}
    \caption{Illustration of optimizing a state diagram with tree topology. The algorithm aims to determine the optimal virtual bond dimensions between sites. This means minimizing the number of vertices (black dots) as the algorithm transitions from the initial tree structure on the left to the optimized structure on the right.}
    \label{fig:tree_before_after}
\end{figure*}

However, the SGE can cause sub-optimal results if the coefficients differ pairwise. Therefore, we apply the bipartite graph algorithm to both the initial \( \Gamma \) and the processed \( \tilde{\Gamma} \), keeping the better result. This ensures that the performance of the previous algorithm is at least matched, with potential improvements in the best-case scenario. 

We additionally adapted a partial pivoting strategy from the method described in \cite{heath2002scientific} and modified it to adhere to the symbolic constraints during the SGE. This can boost the performance of our algorithm. An additional improvement can be achieved by \emph{de-parallelizing} $\Gamma$ before the SGE is performed. The improvement involves \emph{preprocessing} $\Gamma$ once more to detect parallel rows or columns. By \emph{parallel rows or columns}, we refer to those that are either identical or scalar multiples of each other. When detected, we eliminate them by adding a suitably scaled version of one row (or column) to the other to break the dependency. 

Symbolic computations can create challenges in pivoting due to the complexity of managing symbolic expressions. By eliminating parallel rows or columns beforehand, this step simplifies the structure of $\Gamma$, reducing the likelihood of pivoting struggles during the elimination process. This preprocessing step ensures that the matrix is reduced to a more compact form. The full algorithm is given as pseudocode in Alg.~\ref{alg:algorithm}.

\subsubsection{An Example Calculation}\label{sec:example_calc}
Let us illustrate our combined SGE and bipartite graph method approach with the example shown in Fig.~\ref{fig:algorithm}. It is a modified version of the example shown in Fig.~\ref{fig:bipartite}. The matrix corresponding to the initial bipartite graph is given by
\begin{equation}
    \Gamma_h = 
\begin{pmatrix}
3\gamma_{11} & 3\gamma_{12} & -\gamma_{13} & 0 & 2\gamma_{12} \\
0 & 3\gamma_{22} & 0 & 0 & 2\gamma_{22} \\
0 & 3\gamma_{32} & 0 & 0 & 2\gamma_{32} \\
0 & 0 & 2\gamma_{43} & 2\gamma_{44} & 0 \\
0 & 0 & -\gamma_{43} & -\gamma_{44} & 0
\end{pmatrix} .
\end{equation}

If we were to apply the purely bipartite-graph-based algorithm directly, the resulting bond dimension would be $5$. Thus, no compression would take place. However, it becomes evident that some rows or columns can be eliminated, as they are linear combinations of others. Therefore, we first employ the SGE to obtain a reduced-rank matrix \( \Tilde{\Gamma} \) which returns the following three matrices

\begin{align}
A &= 
\begin{array}{c|cccc|}
    & \Tilde{U}_1 & \Tilde{U}_2 & \Tilde{U}_3 &\Tilde{U}_4  \\
    \hline
    U_1 & 1 & 0 & 0 & 0  \\
    U_2 & 0 & 1 & 0 & 0 \\
    U_3 & 0 & 0 & 1 & 0 \\
    U_4 & 0 & 0 & 0 & 1 \\
    U_5 & 0 & 0 & 0 & -1 / 2 \\
    \hline
\end{array}  \\ \nonumber \\
\tilde{\Gamma} &= 
\begin{array}{c|cccc|}
    & \Tilde{V}_1 & \Tilde{V}_2 & \Tilde{V}_3 &\Tilde{V}_4  \\
    \hline
    \Tilde{U}_1 & 3\gamma_{11} & 3\gamma_{12} & -\gamma_{13} & 0 \\
    \Tilde{U}_2 & 0 & 3\gamma_{22} & 0 & 0 \\
    \Tilde{U}_3 & 0 & 3\gamma_{32} & 0 & 0 \\
    \Tilde{U}_4 & 0 & 0 & 2\gamma_{43} & 2\gamma_{44} \\
    \hline
\end{array} \\ \nonumber \\
B &=
\begin{array}{c|ccccc|}
    & \text{V}_1 & \text{V}_2 & \text{V}_3 & \text{V}_4 & \text{V}_5\\
    \hline
    \Tilde{V}_1  & 1 & 0 & 0 & 0 & 0 \\
    \Tilde{V}_2 & 0 & 1 & 0 & 0 & 2 / 3 \\
    \Tilde{V}_3  & 0 & 0 & 1 & 0 & 0 \\
    \Tilde{V}_4  & 0 & 0 & 0 & 1 & 0 \\
    \hline
\end{array}
\end{align}

The details of this computation can be found in Appendix~\ref{app:SGE}. Now we apply a version of the bipartite graph algorithm with respect to $\Tilde{\Gamma}$ as follows:

\begin{figure*}[t!]
    \centering
    \usetikzlibrary{fit, positioning}
    \begin{tikzpicture}[scale=0.8 , transform shape]
    \node (A) at (0,0) {\begin{tikzpicture}[
    regular node/.style={circle, draw, fill=white, inner sep=2pt, minimum size=25pt}
]

\node[regular node] (1) at (0, 0) {1};
\node[regular node] (2) at (0, 1.5) {2};
\node[regular node] (3) at (2, 2.5) {3};
\node[regular node] (4) at (-2, 2.5) {4};
\node[regular node] (5) at (1, -1.5) {5};
\node[regular node] (6) at (2.7, -1.5) {6};
\node[regular node] (7) at (0, -3) {7};
\node[regular node] (8) at (-2, -2) {8};

\draw[orange] (1) -- (2) node[near end, left, label, color=red] {1} node[ near start, right,label, color=blue] {3};
\draw[ darkgreen] (2) -- (3);
\draw[ darkgreen] (2) -- (4);
\draw[ orange] (1) -- (5) node[very near end, left, label, color=red] {2} node[very near start, right,label, color=blue] {4};
\draw[ darkgreen] (5) -- (6);
\draw[ darkgreen] (5) -- (7);
\draw[ pink] (7) -- (8);

\end{tikzpicture}
};
    \node[font=\large] at (-0.4,-4) {(a) BFS traversal from root node $1$};
    \node (B) at(7,0) {\begin{tikzpicture}[
    regular node/.style={circle, draw, fill=white, inner sep=2pt, minimum size=25pt}
]

\node[regular node] (1) at (0, 0) {1};
\node[regular node] (2) at (0, 1.5) {2};
\node[regular node] (3) at (2, 2.5) {3};
\node[regular node] (4) at (-2, 2.5) {4};
\node[regular node] (5) at (1, -1.5) {5};
\node[regular node] (6) at (2.7, -1.5) {6};
\node[regular node] (7) at (0, -3) {7};
\node[regular node] (8) at (-2, -2) {8};

\draw[cyan] (1) -- (2) ;
\draw[ ] (2) -- (3);
\draw[ ] (2) -- (4);
\draw[ ] (1) -- (5);
\draw[ ] (5) -- (6);
\draw[ ] (5) -- (7);
\draw[ ] (7) -- (8);

\draw[dashed, red, rounded corners=10pt, thick] 
    ($(8) + (-0.5,-0.4)$) -- ($(7) + (0.25,-0.7)$) -- 
    ($(5) + (0.2,-0.5)$) -- ($(6) + (0.5,-0.5)$) -- ($(6) + (0.5,0.5)$) -- ($(5) + (0,0.5)$) --
    ($(1) + (0.5,0.5)$) -- ($(1) + (-0.5,0.5)$) -- 
    ($(8) + (-0.5,0.3)$) --cycle;

\draw[dashed, blue, rounded corners=10pt, thick] 
    
    ($(2) + (0.5,-0.5)$) -- ($(3) + (0.5,-0.5)$) -- ($(3) + (0.5,0.5)$)--  ($(4) + (-0.5,0.5)$) -- 
    ($(4) + (-0.5,-0.5)$) -- ($(2) + (-0.5,-0.5)$) -- 
    cycle;

\end{tikzpicture}
};
    \node[font=\large] at (6.8,-4) {(b) Subtrees for split between $1$ and $2$};
    \node (C) at(14,0) {\begin{tikzpicture}[
    regular node/.style={circle, draw, fill=white, inner sep=2pt, minimum size=25pt}    
]

\node[regular node] (1) at (0, 0) {1};
\node[regular node] (2) at (0, 1.5) {2};
\node[regular node] (3) at (2, 2.5) {3};
\node[regular node] (4) at (-2, 2.5) {4};
\node[regular node] (5) at (1, -1.5) {5};
\node[regular node] (6) at (2.7, -1.5) {6};
\node[regular node] (7) at (0, -3) {7};
\node[regular node] (8) at (-2, -2) {8};

\draw[] (1) -- (2);
\draw[ ] (2) -- (3);
\draw[] (2) -- (4);
\draw[ ] (1) -- (5);
\draw[ cyan ] (5) -- (6) ;
\draw[ ] (5) -- (7);
\draw[ ] (7) -- (8);

\draw[dashed, red] (2.7, -1.5) ellipse [x radius=0.6cm, y radius=0.6cm];

\draw[dashed, blue, rounded corners=10pt, thick] 
    ($(8) + (-0.6,-0.5)$) -- ($(7) + (0.1,-0.7)$) -- 
    ($(5) + (0.8,0)$) -- ($(1) + (0.5,0)$) -- 
    ($(2) + (0.5,-0.5)$) -- ($(3) + (0.6,-0.5)$) -- ($(3) + (0.6,0.6)$)--  ($(4) + (-0.6,0.6)$) -- 
    ($(4) + (-0.6,-0.5)$) -- ($(2) + (-0.5,-0.5)$) -- 
    ($(1) + (-0.6,-0.5)$) -- ($(8) + (-0.5,0.5)$) --cycle;

\end{tikzpicture}
};
    \node[font=\large] at (14.2,-4) {(c) Subtrees for split between $5$ and $6$};
    \end{tikzpicture}

    \caption{(a) Tree is traversed in BFS manner but double passing each level. The order of the traverse is first the first level (orange), then the second level (green), and finally the third level (pink). For each layer, we process each vertex site twice. (b) Shows the split for the bond between sites $1$ and $2$. The dashed red and blue outlines show the subtrees on which the $U$ set and $V$ set of nodes corresponding to partial state diagrams are formed, respectively. (c) Another example cut for the bond between sites $5$ and $6$.}
    \label{fig:traverse_order}
\end{figure*}
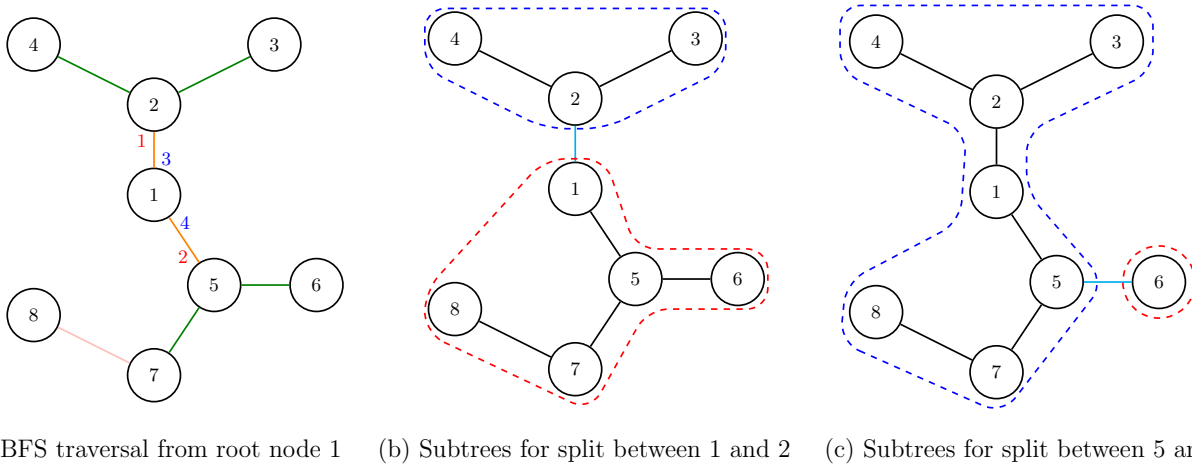

\begin{enumerate}
    \item We interpret the reduced matrix $\tilde{\Gamma}$ as defining two virtual layers of nodes, denoted by $\Tilde{U}$ and $\Tilde{V}$. The new layers are potioned between the old sets $U$ and $V$. This is shown in the first step of Fig.~\ref{fig:algorithm} where the original operator chains are connected via new intermediate virtual nodes $\Tilde{U}_i \in \Tilde{U}$ and $\Tilde{V}_i \in \Tilde{V}$.

    \item We then apply the bipartite graph algorithm to the virtual layers $\Tilde{U}$ and $\Tilde{V}$ to analyze the inner connectivity. Specifically, we compute a minimum vertex cover, as explained in Sec.~\ref{sec:bipartite_method}, on the bipartite graph induced by $\Tilde{U}$ and $\Tilde{V}$ as shown in step 2 of Fig.~\ref{fig:algorithm}.

    \item In the final step, the virtual layers $\Tilde{U}$ and $\Tilde{V}$ are merged back into the original layers $U$ and $V$. During this merging process, the operator chains are reconnected and the updated coefficients are assigned to the corresponding nodes. This reconstruction is depicted in step 3 of Fig.~\ref{fig:algorithm}.
\end{enumerate}

\subsection{Extension to TTNOs}

The bipartite-graph-based algorithm was initially developed to construct MPOs \cite{Ren_2020} and was extended to TTNOs in \cite{10.1063/5.0218773}. Similarly, we will generalize the Algorithm~\ref{alg:algorithm} from MPOs to TTNOs.

We start the optimization with tree-like state diagrams stacked as shown on the left of Fig.~\ref{fig:tree_before_after} (playing the role of the one-dimensional parallel operator chains from before). Due to the tree structure, we can choose a bond and obtain two separated subtrees on either side. Two examples can be found in Fig.~\ref{fig:traverse_order}b-c. From here, the core algorithm stays the same, where we compare and combine partial state diagrams to set up the sets $U$ and $V$. The following optimization is the same as for MPOs.

The order of optimization of the bonds is not a simple left-right sweep through operator chains. Instead, we use a Breadth-First Search (BFS) traversal of the tree structure. We start from an arbitrarily chosen tree root and proceed through its branches. A distinctive aspect of our approach is a \emph{dual traversal} over each level. Rather than processing each bond consecutively, the algorithm completes a full pass through all bonds at a given level before proceeding to the next level. During this dual traversal, each bond is visited twice for implementation purposes. In a typical approach, visiting a bond would involve calculating a non-redundant set for both sides and running the optimization operation. Instead, our method first visits each bond in one level to create a non-redundant set of $U$-subtrees. In the second pass, each bond is revisited to compute a non-redundant set of $V$-subtrees and perform the optimization. This two-phase process is particularly advantageous, as calculating the non-redundant set of $V$-subtrees depends on the previously computed $U$-subtree sets. Consequently, this approach significantly reduces computational complexity. The traversal order for an example tree structure is visually depicted in Fig.~\ref{fig:traverse_order}a. Ultimately, we will have an optimized state diagram easily transformable into a TTNO \cite{milbradt2024state}. An example of an optimized state diagram is depicted on the right in Fig.~\ref{fig:tree_before_after}.

The comparison of partial state diagrams is generally more computationally demanding than partial operator chains. We employ a hashing mechanism to simplify these comparisons to ensure computational efficiency, drastically reducing the computational burden. Hash values are precomputed during tree creation, enabling fast comparisons. For subtrees that do not contain the root, specialized hash functions account for their unique structure, as detailed in App.~\ref{appendix:implementation_details}. This approach, combined with a double-traversed BFS and root selection flexibility, ensures efficient tree structure computation and optimization results.

\section{Experiments and Evaluations}
\label{sec:results}

\begin{figure*}[t]
    \begin{subfigure}[b]{0.5\textwidth}
        \centering
        \includegraphics[height=6cm]{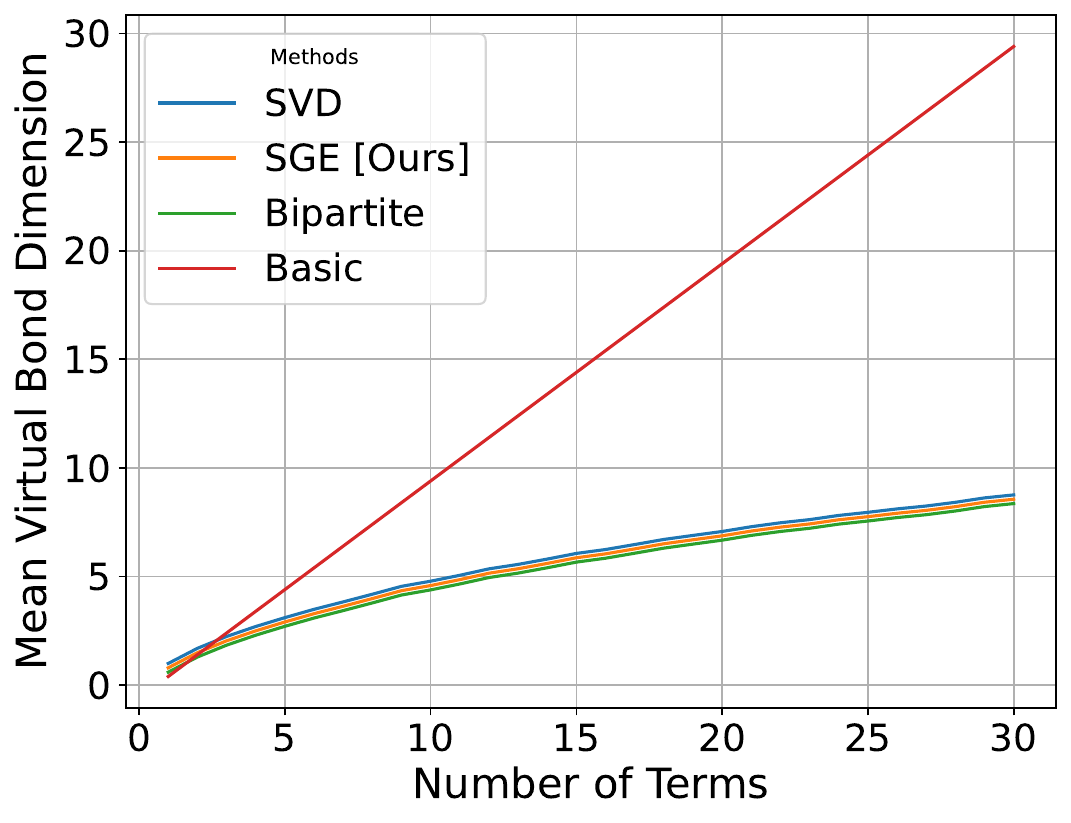}
        \caption{Mean bond dimensions for different methods, with basic configuration and with uniform coefficients}
        \label{fig:uniform_mean_bond_dims_w_basics}
    \end{subfigure}%
    \begin{subfigure}[b]{0.5\textwidth}
        \centering
        \includegraphics[height=6cm]{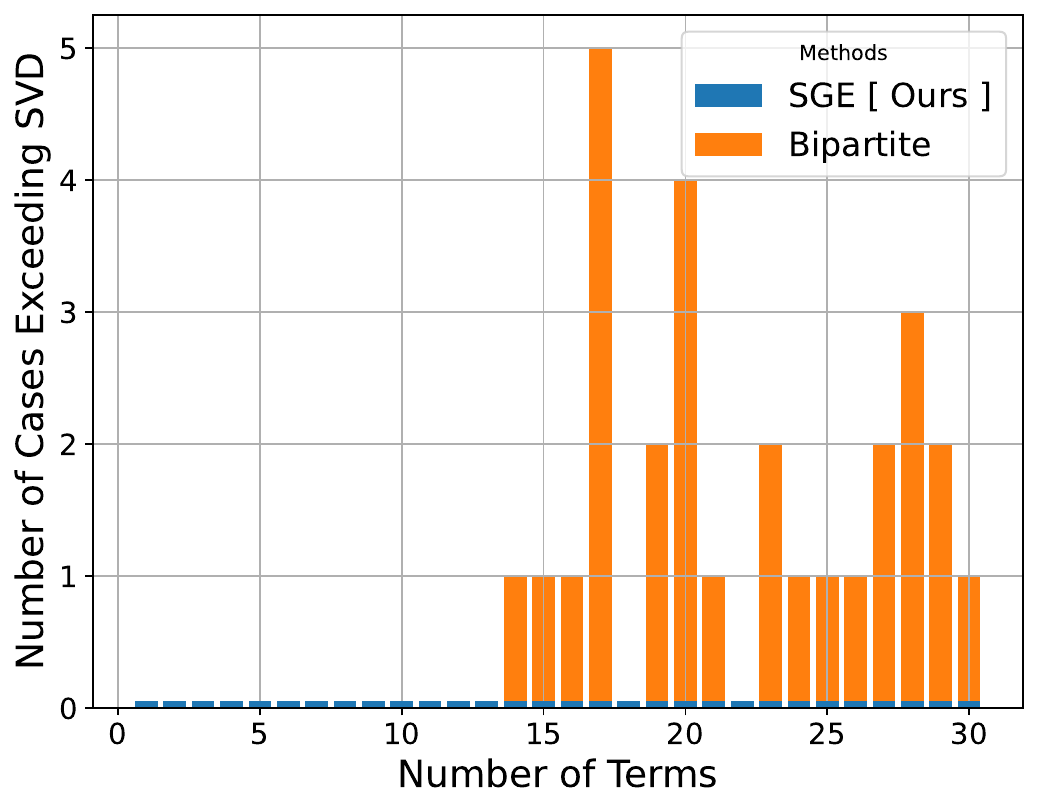}
        \caption{Number of cases where bond dimension is suboptimal compared to SVD, for uniform coefficients}
        \label{fig:uniform_highlight_diff_w}
    \end{subfigure}
    \caption{Numerical results for the random Hamiltonian samples.}
    \label{fig:num_res_random}
\end{figure*}

This chapter presents the numerical experiments to evaluate the proposed algorithms and discusses the obtained results. The experiments test the algorithm's performance and applicability to various quantum systems. Additionally, we test different features for each system, such as the impact of varying tree structures and the effect of diverse coefficient assignments, including uniform, distinct, and discrete values.

To evaluate the efficiency of the constructed tensor networks, we focus on two key metrics. The \emph{mean virtual bond dimension}
\begin{equation}
    D_{\text{mean}} = \sum_{i=1}^M \frac{\dim (\nu_i)}{M}, 
\end{equation}
which indicates the overall efficiency of the optimization, and the \emph{maximum virtual bond dimension}
\begin{equation}
 D_{\max} = \max_{i=1, \dots, M} \left\{ \dim (\nu_i) \right\},
\end{equation}
which highlights the bottleneck of a given tensor network representation, as it directly impacts the computational complexity and resource requirements of simulations. These metrics will be used consistently across the experiments presented in this section.

\subsection{Randomly Generated Hamiltonians}

First, we explore randomly generated Hamiltonians $H$ of the form \eqref{eq:sum_prod_op_repr}. This approach provides a flexible and versatile framework for testing our implementation across various scenarios, as the Hamiltonian parameters can be freely varied to simulate different system configurations. While similar to the randomly generated Hamiltonians in \cite{milbradt2024state}, we can explicitly assign coefficients \(\gamma_k\) in one of the following ways:
\begin{enumerate}[label=(\roman*)]
    \item All coefficients are distinct, $\gamma_k \neq \gamma_{k'}$ for $k \neq k'$.
    \item All coefficients are uniform, $\gamma_k = \gamma$ for all $k$.
    \item Coefficients are partially uniform, which means they are selected from a discrete set of predefined values, $\gamma_k \in S_\gamma$ with $|S_\gamma| < K$, where $K$ is the number of terms in \eqref{eq:sum_prod_op_repr}.
\end{enumerate}

For the local operators \(O_k^{(\ell)}\) used in the products \eqref{eq:sum_prod_op_repr}, we constrain the set of operators to the identity matrix \( \id \) and the three Pauli matrices $X$, $Y$, and $Z$. The primary applications include validating algorithm performance by varying parameters like term count \(K\) and operator complexity, simulating generic quantum systems without specific physical models, and exploring edge cases to identify potential limitations. This approach ensures a comprehensive evaluation of the proposed methods, offering insights into their scalability, robustness, and applicability to real-world systems.

In this setting, we consider random Hamiltonians with $K \in \{1, \dots, 30 \}$ terms, testing $1,000$ distinct Hamiltonians for each $K$, i.e., a total of $30,000$ Hamiltonians for a single set of coefficients. This extensive dataset enables a thorough evaluation of the optimization methods across a diverse range of scenarios.

A fixed tree structure is used consistently throughout the tests, which is the structure shown in the Figures \ref{fig:tree_before_after} and \ref{fig:traverse_order}. By standardizing the tree structure, variations in performance can be attributed solely to the optimization methods, eliminating potential influences from differences in tree topology.

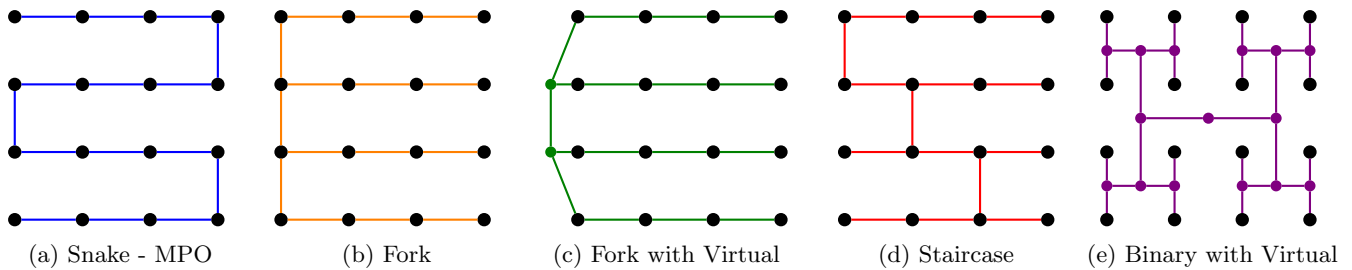
\begin{figure*}[th]
    \centering
    \begin{subfigure}{0.19\linewidth}
        \centering
        \begin{tikzpicture}[every node/.style={circle, draw, fill=black, inner sep=1.5pt}, scale=0.9]

\foreach \x in {0,1,2,3} {
    \foreach \y in {0,1,2,3} {
        \node (node\x\y) at (\x, \y) {};
    }
}
\foreach \y in {0,1,2,3} {
    \draw[thick, blue] (node0\y) -- (node1\y) -- (node2\y) -- (node3\y);
}
\draw[thick, blue] (node30) -- (node31);
\draw[thick, blue] (node32) -- (node33);
\draw[thick, blue] (node01) -- (node02);

\end{tikzpicture}
        \caption{Snake - MPO}
    \end{subfigure}
    \hfill
    \begin{subfigure}{0.19\linewidth}
        \centering
        \begin{tikzpicture}[every node/.style={circle, draw, fill=black, inner sep=1.5pt}, scale=0.9]

\foreach \x in {0,1,2,3} {
    \foreach \y in {0,1,2,3} {
        \node (node\x\y) at (\x, \y) {};
    }
}
\foreach \y in {0,1,2,3} {
    \draw[thick, orange] (node0\y) -- (node1\y) -- (node2\y) -- (node3\y);
}

\draw[thick, orange] (node00) -- (node01) -- (node02) -- (node03);

\end{tikzpicture}
        \caption{Fork}
    \end{subfigure}
    \hfill
    \begin{subfigure}{0.21\linewidth}
        \centering
        
\begin{tikzpicture}[scale=0.9]

\tikzset{regular node/.style={circle, draw, fill=black, inner sep=1.5pt}}
\tikzset{virtual node/.style={circle, draw=darkgreen, fill=darkgreen, inner sep=1.2pt}}

\foreach \x in {0,1,2,3} {
    \foreach \y in {0,1,2,3} {
        \node[regular node] (node\x\y) at (\x, \y) {};
    }
}
\node[virtual node] (node_v1) at (-0.4,1){};
\node[virtual node] (node_v2) at (-0.4,2){};

\foreach \y in {0,1,2,3} {
    \draw[thick, darkgreen] (node0\y) -- (node1\y) -- (node2\y) -- (node3\y);
}

\draw[thick, darkgreen] (node00) -- (node_v1);
\draw[thick, darkgreen] (node_v1) -- (node01);
\draw[thick, darkgreen] (node_v1) -- (node_v2);
\draw[thick, darkgreen] (node_v2) -- (node02);
\draw[thick, darkgreen] (node_v2) -- (node03);

\end{tikzpicture}
        \caption{Fork with Virtual}
    \end{subfigure}
    \hfill
    \begin{subfigure}{0.19\linewidth}
        \centering
        \begin{tikzpicture}[every node/.style={circle, draw, fill=black, inner sep=1.5pt}, scale=0.9]

\foreach \x in {0,1,2,3} {
    \foreach \y in {0,1,2,3} {
        \node (node\x\y) at (\x, \y) {};
    }
}
\foreach \y in {0,1,2,3} {
    \draw[thick, red] (node0\y) -- (node1\y) -- (node2\y) -- (node3\y);
}
\draw[thick, red] (node20) -- (node21);
\draw[thick, red] (node11) -- (node12);
\draw[thick, red] (node02) -- (node03);

\end{tikzpicture}
        \caption{Staircase}
    \end{subfigure}
    \hfill
    \begin{subfigure}{0.19\linewidth}
        \centering
        \begin{tikzpicture}[scale=0.9]

\tikzset{regular node/.style={circle, draw, fill=black, inner sep=1.5pt}}
\tikzset{virtual node/.style={circle, draw=violet, fill=violet, inner sep=1.2pt}}

\foreach \x in {0,1,2,3} {
    \foreach \y in {0,1,2,3} {
        \node[regular node] (node\x\y) at (\x, \y) {};
    }
}

\foreach \x in {0,0.5,1,2,2.5,3} {
    \foreach \y in {0.5,2.5} {
        \pgfmathtruncatemacro{\xName}{\x*10} 
        \pgfmathtruncatemacro{\yName}{\y*10} 
        \node[virtual node] (v_node\xName\yName) at (\x, \y) {};
        \typeout{Creating virtual node at x=\xName, y=\yName}

    }
}

\foreach \x in {0.5,1.5,2.5} {
    \pgfmathtruncatemacro{\xName}{\x*10} 
    
    \typeout{Creating virtual node at x=\xName, y=15}
    \node[virtual node] (v_node\xName15) at (\x, 1.5) {};

}

\foreach \x in {0,1,2,3} {
    \pgfmathtruncatemacro{\xName}{\x*10} 
    \draw[thick, violet] (node\x0) -- (v_node\xName5) -- (node\x1);
    \draw[thick, violet] (node\x2) -- (v_node\xName25) -- (node\x3);
}

\foreach \y in {0.5,2.5} {
    \pgfmathtruncatemacro{\yName}{\y*10} 
    \draw[thick, violet] (v_node0\yName) -- (v_node5\yName) -- (v_node10\yName);
    \draw[thick, violet] (v_node20\yName) -- (v_node25\yName) -- (v_node30\yName);
}

\foreach \x in {5,25} {
    \draw[thick, violet] (v_node\x5) -- (v_node\x15) -- (v_node\x25);
}

\draw[thick, violet] (v_node515) -- (v_node1515) -- (v_node2515);

\end{tikzpicture}
 
        \caption{Binary with Virtual}
    \end{subfigure}
    \caption{Different tensor network topologies spanning a $4\times 4$ square lattice.}
    \label{fig:grid_traverse_methods}
\end{figure*}

The different optimization methods under consideration are as follows:
\begin{itemize}
    \item \textbf{SVD (Singular Value Decomposition)} of the full quantum Hamiltonian: This method provides a baseline for determining optimal bond dimensions but is computationally expensive, scaling as $O(4^L)$ in memory and \(O(4^{3L/2})\) in time for dense matrices with $L$ being the number of physical sites. This makes SVD infeasible for systems larger than $8$ sites in our setup. Consequently, SVD is used as a reference only for small systems, while comparisons among algorithms are conducted for larger systems to ensure computational feasibility.
    \item \textbf{Basic:} The naive construction of the TTNO without any form of compression. It simply stacks the single-term state diagrams, as exemplified in Fig.~\ref{fig:tree_before_after}, and translates the result into a TTNO. This represents the worst-case scenario with maximal bond dimensions.
    \item \textbf{Bipartite Theory:} This approach applies the bipartite graph optimization technique described in Sec.~\ref{sec:bipartite_method} to minimize the bond dimensions.
    \item \textbf{SGE + Bipartite:} This enhanced method combines SGE with the bipartite-graph-based algorithm, as described in Sec.~\ref{sec:SGE}, to preprocess and optimize the TTNO construction.
\end{itemize}
As outlined above, each optimization method was tested with the three distinctly generated coefficient sets.

The results align with theoretical expectations, validating the bipartite graph method's ability to achieve optimal bond dimensions \cite{Ren_2020, 10.1063/5.0218773} in most cases. Our SGE approach consistently reached optimal configurations across all tested cases, including those with partially uniform and distinct coefficient sets. While partially uniform coefficients could theoretically introduce non-optimality in the plain bipartite graph method, this was not observed, likely due to a sufficient number of unique terms and the randomness of the experiment set, avoiding problematic edge cases. As the distinct and partially uniform sets had all methods performed equally, we did not plot them.

The interesting case is the uniform coefficients scenario, for which the results are plotted in Fig.~\ref{fig:num_res_random}. This scenario highlights the limitations of the vanilla bipartite graph method and the improvements that SGE introduced. As shown in Fig.~\ref{fig:uniform_mean_bond_dims_w_basics}, the naïve construction yields significantly higher mean bond dimensions, emphasizing the efficiency of the optimized methods. While the mean bond dimensions for the bipartite graph and SGE-enhanced approaches appear similar, the detailed analysis in Fig.~\ref{fig:uniform_highlight_diff_w} reveals key differences. The standalone bipartite graph method shows increasing deviations from SVD results as system complexity grows. In contrast, the SGE-enhanced approach consistently matches the optimal bond dimensions achieved by SVD, demonstrating its effectiveness in addressing the limitations of the bipartite graph method, particularly for more complex uniform coefficient cases.

\subsection{Effective Lattice Hamiltonian}

Even though the previous Hamiltonians capture both short and long-range interactions within a quantum system, an additional feature we aim to explore is the scenario where coefficients are shared across different terms. To simplify the system under consideration while retaining both local and long-range relations, we adopt the following Hamiltonian, where the interaction strength decays with distance,
\begin{equation}\label{eq:lattice_ham}
H = \sum_{i \neq j} \frac{J}{\|i - j\|} X_i X_j + \sum_i g Z_i.
\end{equation}
$X$ and $Z$ are once more the Pauli-matrices, $J \in \R$ is the interaction coupling strength, and $g \in \R$ the external magnetic field strength. For the distance between two sites \(i\) and \(j\), \(\|i - j\|\), we chose the Manhattan distance. The Manhattan distance is defined as the sum of the absolute differences of their coordinates and the total "grid-based" separation between two points on a 2D grid. Due to this selection, the factors of the interaction term will all be rational multiples of $J$. The sites on which the Hamiltonian $H$ is defined are arranged in an $L \times L$ grid.

This second set of experiments highlights a specific use case of our algorithm, focusing on the effective lattice Hamiltonian given in \eqref{eq:lattice_ham}, featuring both local and long-range interactions. We evaluated the influence of tree structures on the optimization process only with the SGE-enhanced approach. The algorithm's ability to handle decaying interaction strengths was tested using a two-dimensional lattice model. We employed multiple methods to span the lattice grid with a tree, allowing for a detailed comparison of tree configurations and their impact on bond dimension optimization within the framework of our algorithm. The different tree structures used are:

\begin{figure*}[t!]
    \begin{subfigure}[b]{0.49\textwidth}
        \centering
        \includegraphics[width=\linewidth]{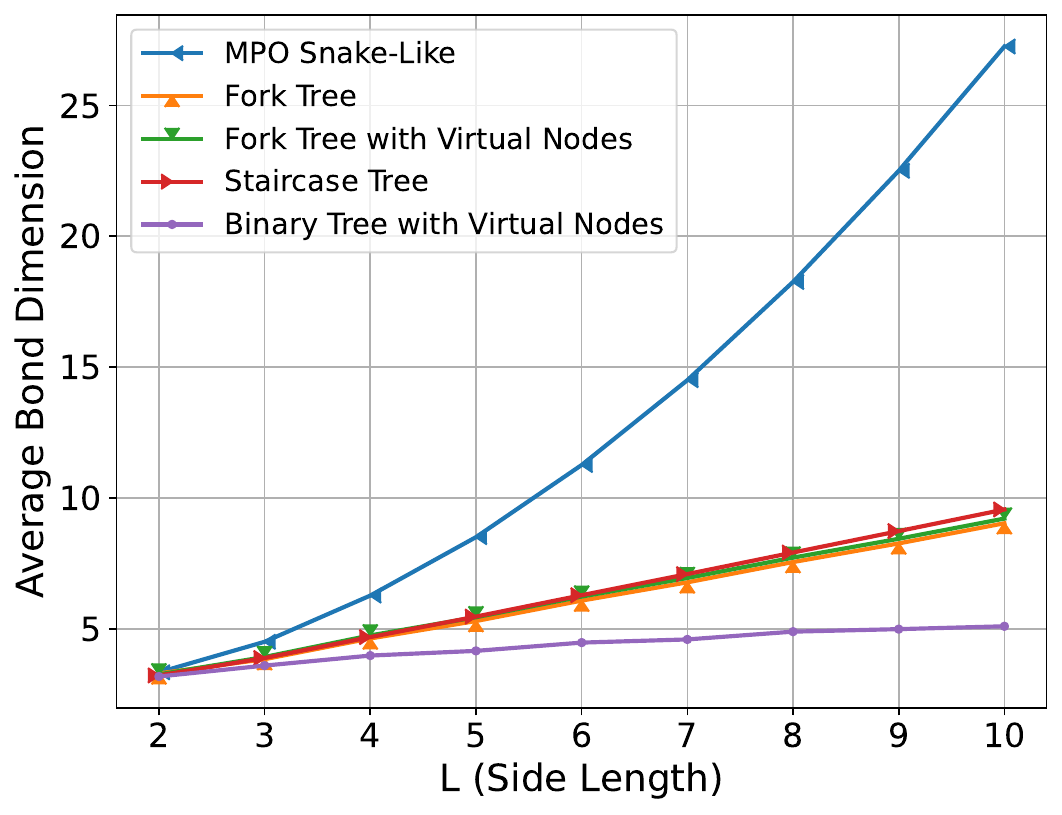}
        \caption{Average bond dimensions for different methods, with basic configuration}
        \label{fig:mean_grid}
    \end{subfigure}
    \begin{subfigure}[b]{0.5\textwidth}
        \centering
        \includegraphics[width=\linewidth]{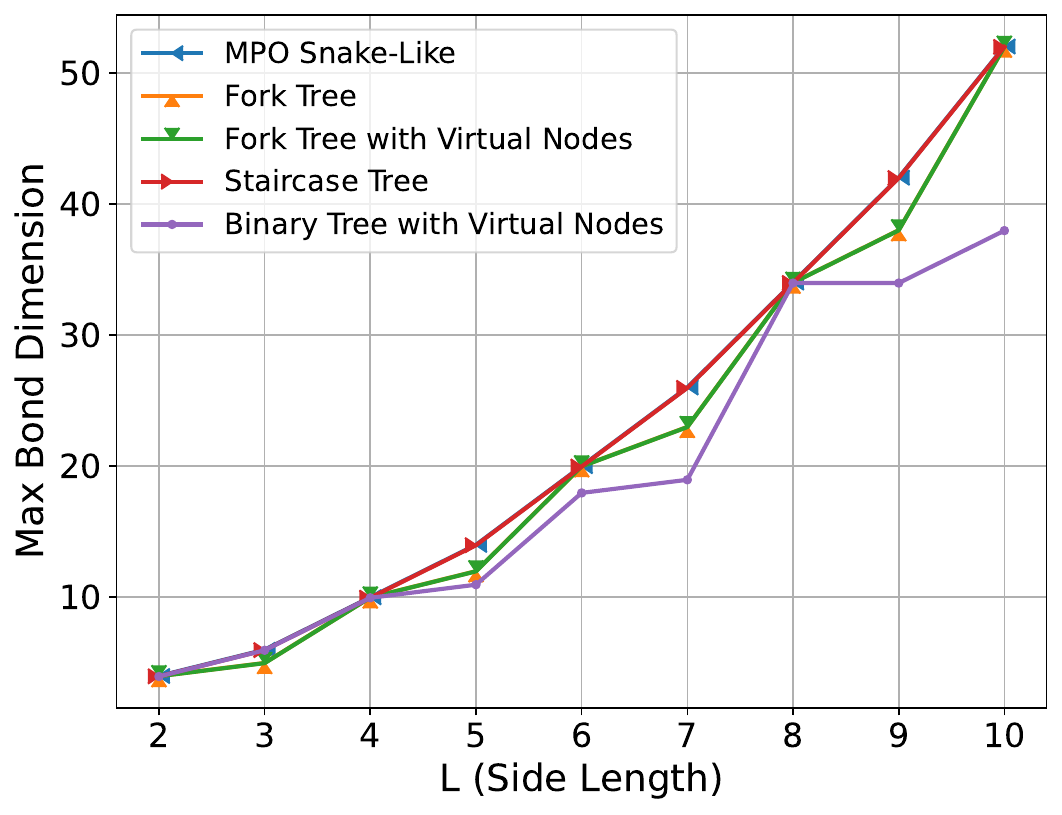}
        \caption{Maximum bond dimensions for different methods, with basic configuration}
        \label{fig:max_grid}
    \end{subfigure}
    \caption{Numerical results for the construction of the lattice Hamiltonian \eqref{eq:lattice_ham} using SGE for the different tree structures given in Fig.~\ref{fig:grid_traverse_methods}.}
    \label{fig:lattice_results}
\end{figure*}

\begin{itemize}
    \item \textbf{Snake - MPO}: This structure runs through the lattice in a snake-like pattern, effectively creating an MPO. This and similar approaches are commonly used to simulate 2D systems with tensor networks \cite{Xiang2001, Cataldi2021}.

    \item \textbf{Fork Layout}: A tree structure where the first node of each line is connected directly to the other first nodes, creating a layered hierarchy across the grid \cite{Bauernfeind2017}.

    \item \textbf{Fork with Virtual Nodes}: A variation of the Fork tree structure to make it adhere to the T3NS formalism \cite{Gunst2018}. The connections between the first nodes are established through additional virtual nodes, providing an intermediary layer. 

    \item \textbf{Staircase}: A more balanced tree structure that traverses the grid along the diagonal, creating a hierarchical tree configuration stemming from the diagonal connections.
    
    \item \textbf{Binary}: A binary tree structure where only the leaves have physical legs, and the relational hierarchy is provided by virtual nodes \cite{Kloss2020}.
\end{itemize}

They are also depicted in Fig.~\ref{fig:grid_traverse_methods} for a $4\times 4$ square lattice. As in the previous experiments, we evaluate the performance using the mean and maximum virtual bond dimensions which quantify the overall efficiency and computational bottlenecks of the tensor network representations.

The results are shown in Fig.~\ref{fig:lattice_results}. Fig.~\ref{fig:mean_grid} visualizes the mean virtual bond dimensions as a function of lattice length \( L \); note that the total number of sites is \( L^2 \). The snake-like MPO exhibits a rapid increase in bond dimensions with \( L \), reflecting its inefficiency in capturing the long-range interactions due to its linear topology. Other methods demonstrate significantly better scalability compared to the snake-like shape. These topologies effectively balance local and long-range correlations, leading to much slower growth in bond dimensions as the system size increases. This improvement underscores the advantages of tree structures, particularly for systems with long-range interactions. The binary tree, in particular, achieves the lowest mean bond dimension growth, benefiting from its hierarchical structure and the additional virtual nodes.

The graph in Fig.~\ref{fig:max_grid} illustrates each structure's maximum virtual bond dimension across varying grid sizes $L$. The maximum bond dimension is more relevant for evaluating the computational cost of algorithms like DMRG using the MPO or TTNO. As \( L \) increases, the snake-like topology exhibits the largest maximum bond dimensions, highlighting its inefficiency in capturing long-range interactions due to its linear topology. Interestingly, the staircase tree performs similarly, indicating its structure still behaves like a linear layout despite its intended hierarchical design. In contrast, the fork trees with and without virtual nodes show better scalability with smaller maximum bond dimensions. These structures effectively balance local and global correlations, though the introduction of virtual nodes does not appear to significantly impact the maximum bond dimensions observed in the system.

\begin{figure*}[ht]
    \begin{subfigure}[b]{0.49\textwidth}
        \raisebox{0.5cm}{
        \begin{tikzpicture}[every node/.style={circle, draw, inner sep=3pt}]
\def\cavcolour{pink}
\def\molcolour{green}
\def\bathcolour{orange}
\def\horizdist{0.75}
\def\vertdist{1}
\def\physdist{0.4}

\path[use as bounding box](-1,\physdist+0.1) rectangle (10*\horizdist+0.2, -3*\vertdist+0.5);

\node[fill=\bathcolour] (MB11) at (0,0){};
\draw (MB11) -- (0,\physdist);
\node[fill=\bathcolour] (MB12) at (\horizdist,0){};
\draw (MB12) -- (\horizdist,\physdist);
\node[fill=\bathcolour] (MB13) at (2*\horizdist,0){};
\draw (MB13) -- (2*\horizdist,\physdist);
\node[fill=\molcolour] (M11) at (3*\horizdist,0){};
\draw (M11) -- (3*\horizdist,\physdist);
\node[fill=\molcolour] (M12) at (4*\horizdist,0){};
\draw (M12) -- (4*\horizdist,\physdist);
\node[fill=\bathcolour] (MB21) at (5*\horizdist,0){};
\draw (MB21) -- (5*\horizdist,\physdist);
\node[fill=\bathcolour] (MB22) at (6*\horizdist,0){};
\draw (MB22) -- (6*\horizdist,\physdist);
\node[fill=\bathcolour] (MB23) at (7*\horizdist,0){};
\draw (MB23) -- (7*\horizdist,\physdist);
\node[fill=\molcolour] (M21) at (8*\horizdist,0){};
\draw (M21) -- (8*\horizdist,\physdist);
\node[fill=\molcolour] (M22) at (9*\horizdist,0){};
\draw (M22) -- (9*\horizdist,\physdist);

\begin{scope}[shift={(-2.5*\horizdist,0)}]
    \node[fill=\cavcolour] (C1) at (5*\horizdist,-1*\vertdist){};
    \draw (C1) -- (5*\horizdist,\physdist-1*\vertdist);
    \node[fill=\cavcolour] (C2) at (6*\horizdist,-1*\vertdist){};
    \draw (C2) -- (6*\horizdist,\physdist-1*\vertdist);
    \node[fill=\bathcolour] (CB1) at (7*\horizdist,-1*\vertdist){};
    \draw (CB1) -- (7*\horizdist,\physdist-1*\vertdist);
    \node[fill=\bathcolour] (CB2) at (8*\horizdist,-1*\vertdist){};
    \draw (CB2) -- (8*\horizdist,\physdist-1*\vertdist);
    \node[fill=\bathcolour] (CB3) at (9*\horizdist,-1*\vertdist){};
    \draw (CB3) -- (9*\horizdist,\physdist-1*\vertdist);
\end{scope}

\begin{scope}[shift={(0,-2*\vertdist)}]
    \node[fill=\molcolour] (M31) at (0,0){};
    \draw (M31) -- (0,\physdist);
    \node[fill=\molcolour] (M32) at (\horizdist,0){};
    \draw (M32) -- (\horizdist,\physdist);
    \node[fill=\bathcolour] (MB31) at (2*\horizdist,0){};
    \draw (MB31) -- (2*\horizdist,\physdist);
    \node[fill=\bathcolour] (MB32) at (3*\horizdist,0){};
    \draw (MB32) -- (3*\horizdist,\physdist);
    \node[fill=\bathcolour] (MB33) at (4*\horizdist,0){};
    \draw (MB33) -- (4*\horizdist,\physdist);
    \node[fill=\molcolour] (M41) at (5*\horizdist,0){};
    \draw (M41) -- (5*\horizdist,\physdist);
    \node[fill=\molcolour] (M42) at (6*\horizdist,0){};
    \draw (M42) -- (6*\horizdist,\physdist);
    \node[fill=\bathcolour] (MB41) at (7*\horizdist,0){};
    \draw (MB41) -- (7*\horizdist,\physdist);
    \node[fill=\bathcolour] (MB42) at (8*\horizdist,0){};
    \draw (MB42) -- (8*\horizdist,\physdist);
    \node[fill=\bathcolour] (MB43) at (9*\horizdist,0){};
    \draw (MB43) -- (9*\horizdist,\physdist);
\end{scope}

\draw (MB11) -- (MB12) -- (MB13) -- (M11) -- (M12)  -- (MB21) -- (MB22) -- (MB23) -- (M21) -- (M22);
\draw (M22) to[out=0,in=0,looseness=1.5] (CB3);
\draw (CB3) -- (CB2) -- (CB1) -- (C2) -- (C1);
\draw (C1) to[in=180,out=180,looseness=1.5] (M31);
\draw (M31) -- (M32) -- (MB31) -- (MB32) -- (MB33) -- (M41) -- (M42) -- (MB41) -- (MB42) -- (MB43);
\end{tikzpicture}
        }
        \caption{The MPS structure that is used as ansatz to solve the HEOM.}
        \label{fig:heom_mps}
    \end{subfigure}
    \begin{subfigure}[b]{0.5\textwidth}
        \centering
        \begin{tikzpicture}[every node/.style={circle, draw, inner sep=3pt}]
\def\cavcolour{pink}
\def\molcolour{green}
\def\virtcolour{yellow}
\def\bathcolour{orange}
\def\horizdist{0.5}
\def\vertdist{1}
\def\physdist{0.4}
\def\nonvirtdist{1 / 1.5}
\def\moldist{\vertdist / 2}

\node[fill=\virtcolour] (V1) at (0,0) {};
\node[fill=\virtcolour] (V21) at (-\horizdist,\vertdist) {};
\node[fill=\virtcolour] (V22) at (-\horizdist,-\vertdist) {};
\draw (V21) -- (V1) -- (V22);

\node[fill=\cavcolour] (C1) at (\nonvirtdist,0) {};
\draw (C1) -- (\nonvirtdist,\physdist);
\node[fill=\cavcolour] (C2) at (2*\nonvirtdist,0) {};
\draw (C2) -- (2*\nonvirtdist,\physdist);
\node[fill=\bathcolour] (CB1) at (3*\nonvirtdist,0) {};
\draw (CB1) -- (3*\nonvirtdist,\physdist);
\node[fill=\bathcolour] (CB2) at (4*\nonvirtdist,0) {};
\draw (CB2) -- (4*\nonvirtdist,\physdist);
\node[fill=\bathcolour] (CB3) at (5*\nonvirtdist,0) {};
\draw (CB3) -- (5*\nonvirtdist,\physdist);
\draw (V1) -- (C1) -- (C2) -- (CB1) -- (CB2) -- (CB3);

\node[fill=\molcolour] (M11) at (-2*\nonvirtdist,\vertdist+\moldist) {};
\draw (M11) -- (-2*\nonvirtdist,\vertdist+\moldist+\physdist);
\node[fill=\molcolour] (M12) at (-3*\nonvirtdist,\vertdist+\moldist) {};
\draw (M12) -- (-3*\nonvirtdist,\vertdist+\moldist+\physdist);
\node[fill=\molcolour] (M21) at (-2*\nonvirtdist,\vertdist-\moldist) {};
\draw (M21) -- (-2*\nonvirtdist,\vertdist-\moldist+\physdist);
\node[fill=\molcolour] (M22) at (-3*\nonvirtdist,\vertdist-\moldist) {};
\draw (M22) -- (-3*\nonvirtdist,\vertdist-\moldist+\physdist);
\node[fill=\molcolour] (M31) at (-2*\nonvirtdist,-\vertdist+\moldist) {};
\draw (M31) -- (-2*\nonvirtdist,-\vertdist+\moldist+\physdist);
\node[fill=\molcolour] (M32) at (-3*\nonvirtdist,-\vertdist+\moldist) {};
\draw (M32) -- (-3*\nonvirtdist,-\vertdist+\moldist+\physdist);
\node[fill=\molcolour] (M41) at (-2*\nonvirtdist,-\vertdist-\moldist) {};
\draw (M41) -- (-2*\nonvirtdist,-\vertdist-\moldist+\physdist);
\node[fill=\molcolour] (M42) at (-3*\nonvirtdist,-\vertdist-\moldist) {};
\draw (M42) -- (-3*\nonvirtdist,-\vertdist-\moldist+\physdist);
\node[fill=\bathcolour] (MB11) at (-4*\nonvirtdist,\vertdist+\moldist) {};
\draw (MB11) -- (-4*\nonvirtdist,\vertdist+\moldist+\physdist);
\node[fill=\bathcolour] (MB12) at (-5*\nonvirtdist,\vertdist+\moldist) {};
\draw (MB12) -- (-5*\nonvirtdist,\vertdist+\moldist+\physdist);
\node[fill=\bathcolour] (MB13) at (-6*\nonvirtdist,\vertdist+\moldist) {};
\draw (MB13) -- (-6*\nonvirtdist,\vertdist+\moldist+\physdist);
\draw (V21) -- (M11) -- (M12) -- (MB11) -- (MB12) -- (MB13);
\node[fill=\bathcolour] (MB21) at (-4*\nonvirtdist,\vertdist-\moldist) {};
\draw (MB21) -- (-4*\nonvirtdist,\vertdist-\moldist+\physdist);
\node[fill=\bathcolour] (MB22) at (-5*\nonvirtdist,\vertdist-\moldist) {};
\draw (MB22) -- (-5*\nonvirtdist,\vertdist-\moldist+\physdist);
\node[fill=\bathcolour] (MB23) at (-6*\nonvirtdist,\vertdist-\moldist) {};
\draw (MB23) -- (-6*\nonvirtdist,\vertdist-\moldist+\physdist);
\draw (V21) -- (M21) -- (M22) -- (MB21) -- (MB22) -- (MB23);
\node[fill=\bathcolour] (MB31) at (-4*\nonvirtdist,-\vertdist+\moldist) {};
\draw (MB31) -- (-4*\nonvirtdist,-\vertdist+\moldist+\physdist);
\node[fill=\bathcolour] (MB32) at (-5*\nonvirtdist,-\vertdist+\moldist) {};
\draw (MB32) -- (-5*\nonvirtdist,-\vertdist+\moldist+\physdist);
\node[fill=\bathcolour] (MB33) at (-6*\nonvirtdist,-\vertdist+\moldist) {};
\draw (MB33) -- (-6*\nonvirtdist,-\vertdist+\moldist+\physdist);
\draw (V22) -- (M31) -- (M32) -- (MB31) -- (MB32) -- (MB33);
\node[fill=\bathcolour] (MB41) at (-4*\nonvirtdist,-\vertdist-\moldist) {};
\draw (MB41) -- (-4*\nonvirtdist,-\vertdist-\moldist+\physdist);
\node[fill=\bathcolour] (MB42) at (-5*\nonvirtdist,-\vertdist-\moldist) {};
\draw (MB42) -- (-5*\nonvirtdist,-\vertdist-\moldist+\physdist);
\node[fill=\bathcolour] (MB43) at (-6*\nonvirtdist,-\vertdist-\moldist) {};
\draw (MB43) -- (-6*\nonvirtdist,-\vertdist-\moldist+\physdist);
\draw (V22) -- (M41) -- (M42) -- (MB41) -- (MB42) -- (MB43);

\end{tikzpicture}
        \caption{The TTN structure that is used as ansatz to solve the HEOM.}
        \label{fig:heom_ttn}
    \end{subfigure}
    \caption{The tensor network structures used as ans\"atze to solve the HEOM for $\mathcal{N} = 4$. The green sites correspond to the molecules, while the pink nodes are the cavity sites. The bath nodes are colored orange, while any virtual nodes are colored yellow.}
    \label{fig:heom_tns}
\end{figure*}
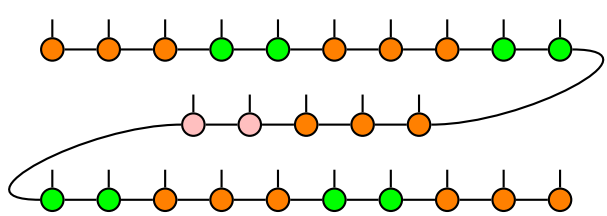
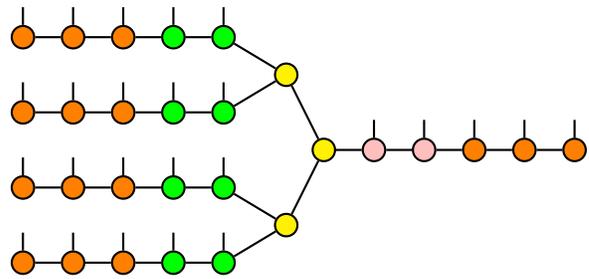

The binary tree with virtual nodes achieves the lowest maximum bond dimensions across all tested system sizes. Its hierarchical organization and use of virtual nodes enable an effective distribution of correlations.
Surprisingly, all topologies exhibit the same maximum bond dimensions for \( L = 2, 4, \) and \( 8 \). This behavior stems from the intrinsic nature of the binary tree structure. Whenever the number of nodes reaches a power of $2$, a new level must be added to the tree, introducing a sudden increase in bond dimensions. This structural adjustment temporarily reduces the efficiency of the binary tree at these specific lattice sizes.

\subsection{Application: Molecules in a Cavity}\label{sec:appl_molecules}

We use the model of a cavity filled with molecules in a solvent to test our methodology on a realistic application. We will describe this model only on a very abstract level; refer to \cite{Ke2024twomol} for more details. Any open quantum system can be described by a Hamiltonian of the form
\begin{equation}
    H_{\text{tot}} = H_S + H_I + H_E,
\end{equation}
where $H_S$ is the Hamiltonian of the main system of interest $S$, in this case the molecules and cavity, $H_I$ the interaction of the system $S$ with the environment $E$, and $H_E$ is the environment's Hamiltonian, in this case the solvent and the cavity environment. For $\mathcal{M}$ molecules, the system Hamiltonian is given by
\begin{equation}\label{eq:system_ham}
\begin{split}
    H_S = & H_{\text{cav}}^{(0)} + \sum_{i=1}^\mathcal{M} \left( \eta_i A^{(0)} \otimes A^{(i)} + H_{\text{mol}}^{(i)} \right) \\
    & + \sum_{i<j} \Delta_{ij} B^{(i)} \otimes B^{(j)},
\end{split}
\end{equation}
where $H_{\text{cav}}$ and $H_{\text{mol}}$ are the Hamiltonians acting on the cavity and a single molecule respectively, $\eta_i$ is the interaction strength between the cavity and the $i$th molecule and $\Delta_{ij}$ is the interaction strength between two molecules. An operator $O^{(k)}$ acts on the cavity for $k=0$ and on the respective molecule for $k \in \{1, \dots, \mathcal{M}\}$. For the exact form of the operators $H_{\text{cav}}$, $H_{\text{mol}}$, $A$, $\tilde{A}$, and $B$ refer to \cite{Ke2024twomol}.

Such a system can be simulated using the hierarchical equations of motion (HEOM)\cite{Tanimura1989,Ishizaki2005}; refer to \cite{Tanimura2020} for a recent review of the method. In the HEOM approach the quantum dynamics are described by an infinite set of coupled differential equations. These can be recast into an effective Schr\"{o}dinger equation describing the dynamics of a state of the main system $S$ given as a density matrix $\rho$. The Schr\"{o}dinger equation is gouverned by the effective super-Hamiltonian
\begin{equation}\label{eq:heom_eff_ham}
\begin{split}
     \mathcal{H} & = \hat{H}_S - \check{H}_S -i \sum_{\alpha=0}^{M} \sum_{\beta=1}^{\mathcal{N}} \gamma_{\alpha\beta} b^\dagger_{\alpha\beta} b_{\alpha\beta} \\
     & + \sum_{\alpha=0}^{M} \sum_{\beta=1}^{\mathcal{N}}\left( \lambda_{\alpha\beta} \hat{L}_\alpha - \lambda_{\alpha\beta}^* \check{L}^\dagger_\alpha \right) b_{\alpha\beta} \\
     & + \sum_{\alpha=0}^{M} \sum_{\beta=1}^{\mathcal{N}}\left( \chi_{\alpha\beta} \hat{L}_\alpha - \chi_{\alpha\beta}^* \check{L}^\dagger_\alpha \right) b_{\alpha\beta}^\dagger, \\
\end{split}
\end{equation}
where $\gamma_{\alpha\beta}$, $\chi_{\alpha\beta}$, and $\lambda_{\alpha\beta}$ are complex constant coefficients and $L$ are operators describing the interaction with the of the main system $S$ with the environment. $b_{\alpha\beta}$ and $b_{\alpha\beta}^\dagger$ are bosonic annihilation and creation operators. Furthermore, we define
\begin{equation}
    \hat{O} = O\rho \quad \text{and} \quad \check{O} = \rho \, O.
\end{equation}
For more details on the derivation refer to Appendix~\ref{appendix:der_of_sham}. The super-Hamiltonian $\mathcal{H}$ is well suited to be used with a tensor network ansatz to solve the HEOM, as was done for MPS \cite{Mangaud2023} and TTN \cite{Ke2023}.

\begin{figure*}[t]
    \begin{subfigure}[b]{0.49\textwidth}
        \includegraphics[width=\textwidth]{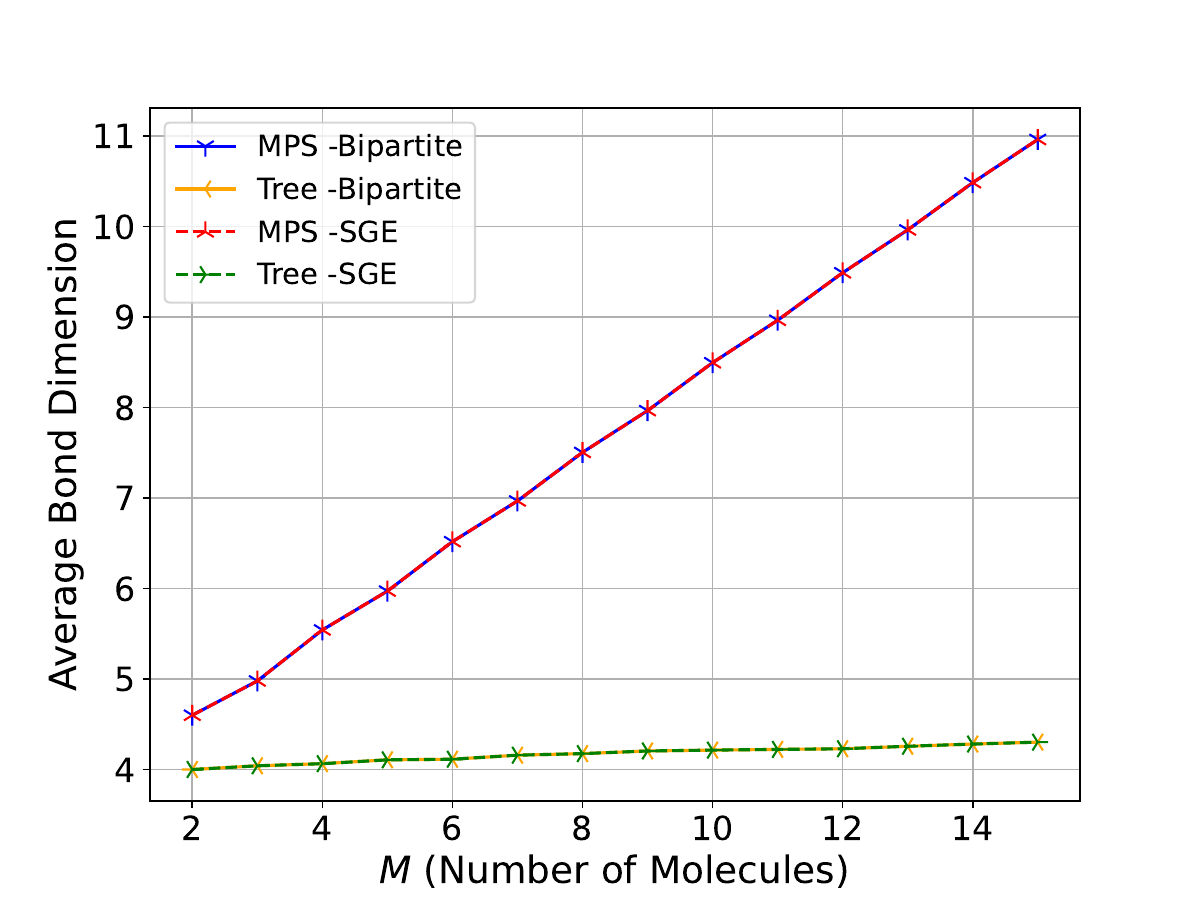}
        \caption{}
        \label{fig:heom_het_avg}
    \end{subfigure}
    \begin{subfigure}[b]{0.5\textwidth}
        \centering
        \includegraphics[width=\textwidth]{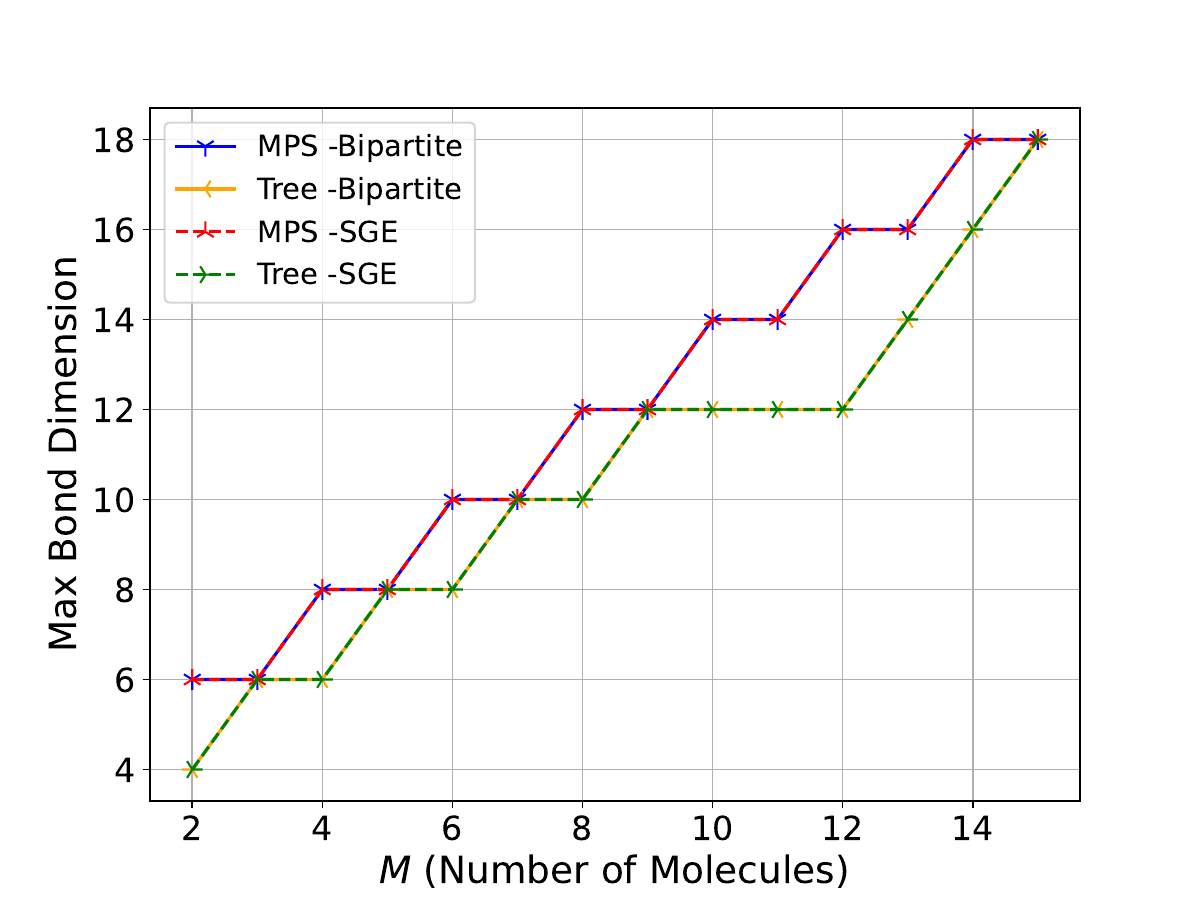}
        \caption{}
        \label{fig:heom_het_max}
    \end{subfigure}
    \begin{subfigure}[b]{0.49\textwidth}
        \includegraphics[width=\textwidth]{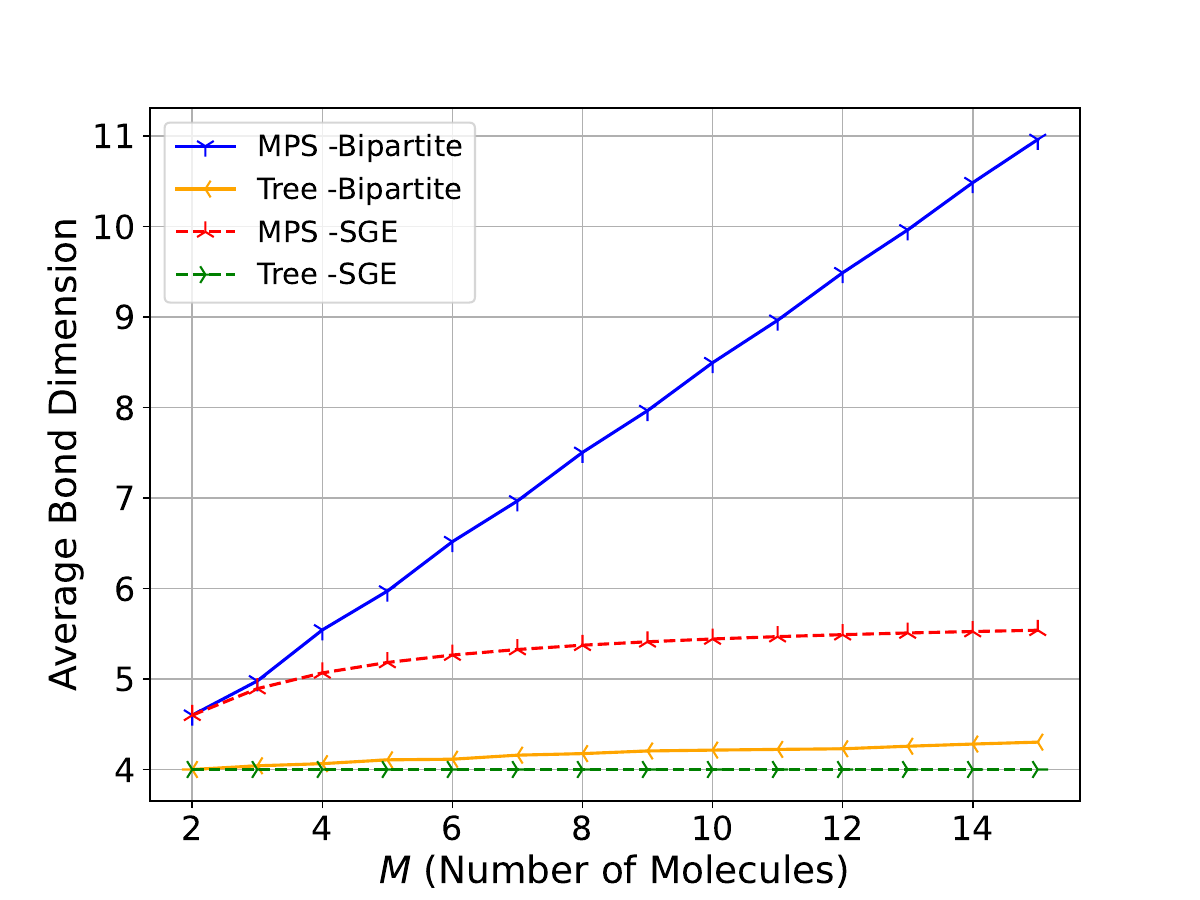}
        \caption{}
        \label{fig:heom_hom_avg}
    \end{subfigure}
    \begin{subfigure}[b]{0.5\textwidth}
        \centering
        \includegraphics[width=\textwidth]{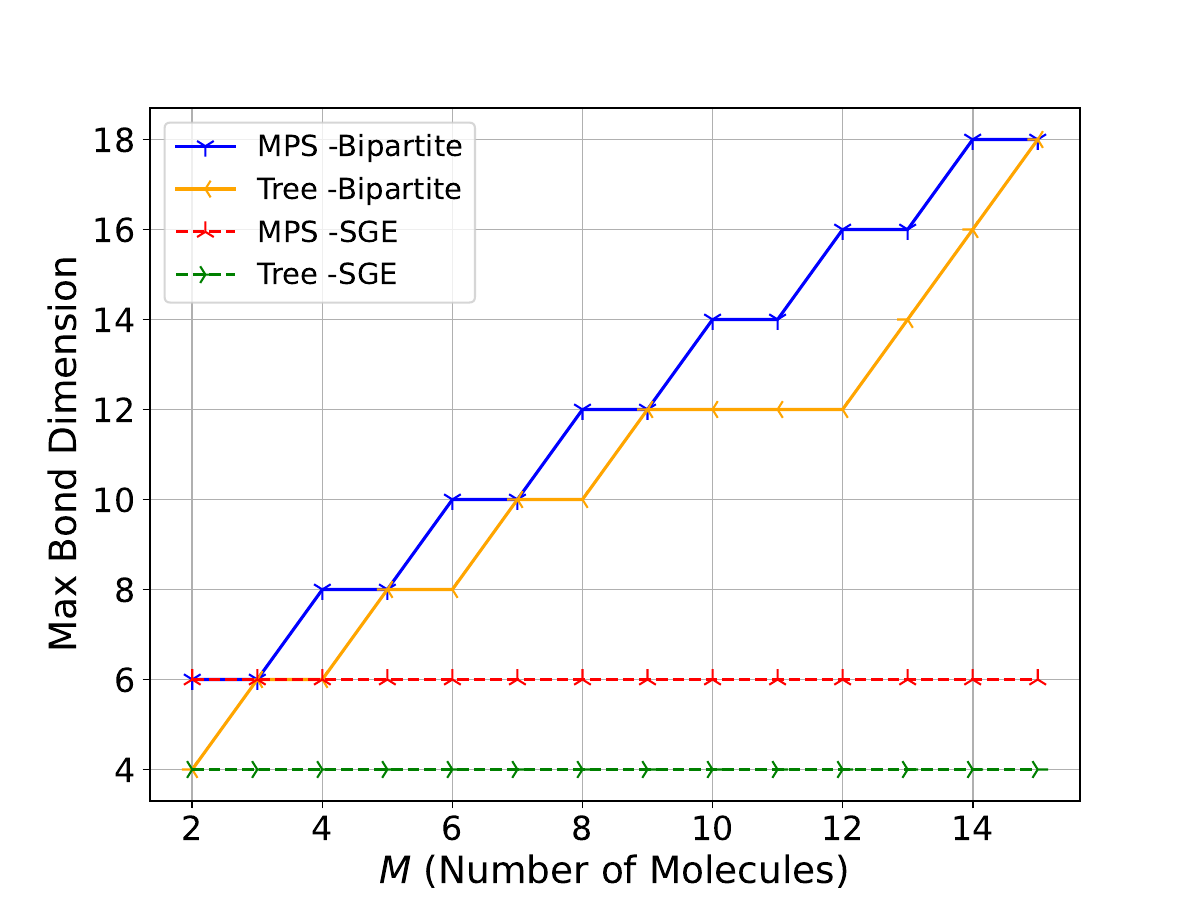}
        \caption{}
        \label{fig:heom_hom_max}
    \end{subfigure}
    \caption{The bond dimensions obtained from running the bipartite and the symbolic Gaussian elimination algorithms for the super-Hamiltonian \eqref{eq:heom_eff_ham} on the tensor network structures shown in Fig.~\ref{fig:heom_tns}, but for $\mathcal{N}=10$. The upper plots show the results for the heterogenous Hamiltonian, while the lower plots show the results for a homogenous Hamiltonian. }
    \label{fig:heom_plots}
\end{figure*}

We will now show the capability of our method by constructing the MPO and TTNO for the super-Hamiltonian $\mathcal{H}$ in Eq.~\eqref{eq:heom_eff_ham}. The MPS structure is shown in Fig.~\ref{fig:heom_mps} and adapted from \cite{Ke2024twomol}. The cavity and each molecule are represented by two sites each, one for the input and one for the output local physical degree of freedom each. The sites representing the hierarchy indices are attached directly to one of the two respective physical sites. The chains representing the molecules are separated into two groups and attached to each side of the cavity and bath sites. This setup aims to minimize the bond dimension increase of the tensor network state during the dynamics governed by $\mathcal{H}$ \cite{Ke2024twomol}. The TTN structure is shown in Fig.~\ref{fig:heom_ttn}. The molecule and cavity are again represented by two sites with the corresponding hierarchy indices attached as chains. The tree structure is almost that of a binary tree, to whose leaves the molecular chains are attached. However, the cavity tensor chain is attached to the root of the binary tree. This is the TTN structure used in \cite{Ke2025} to evaluate the dynamics gouverned by $\mathcal{H}$. We use $\mathcal{N}=10$ for our construction and consider two cases. We call the first case heterogenous, where the factors $\eta_i$, $\Delta_{ij}$, $\gamma_{i\beta}$, $\lambda_{i\beta}$, and $\chi_{i\beta}$ in \eqref{eq:system_ham} and \eqref{eq:heom_eff_ham} differ for different molecular indices $i,j$. The other case is the homogenous case where we assume that all molecules have the same interaction with the cavity, each other, and the environment. Thus, the factors in \eqref{eq:system_ham} and \eqref{eq:heom_eff_ham} are constant with regard to the molecular index.

The bond dimensions obtained from the constructions are plotted in Fig.~\ref{fig:heom_plots} with respect to the number of molecules $M$. The change for differing $\mathcal{N}$ values is not as significant. We compared the MPS and TTN structure in every plot for the bare bipartite construction method and our advanced method based on SGE. In every case, the TTN structure outperforms the MPS structure. This improvement also applies to the actual simulation results, as shown in \cite{Ke2025}. Comparing the two construction methods, we can see no difference for the heterogenous case shown in Fig.~\ref{fig:heom_het_avg} and Fig.~\ref{fig:heom_het_max}. This fits the optimality proof in \cite{Ren_2020} for non-repeating factors and is to be expected. However, the proof's assumption is not true in the homogenous case. Accordingly, we see a significant improvement for the MPS and TTN structures when using SGE as shown in Fig.~\ref{fig:heom_hom_avg} and Fig.~\ref{fig:heom_hom_max}. The scaling of the average bond dimensions goes from linear to sublinear for the MPS structure and improves from linear to constant for the TTNS. The maximum bond dimension scaling becomes constant when using the SGE construction for the MPS and TTN structure while increasing for the bipartite construction method. These numerics show that the TTNO bond dimension can significantly improve using our SGE method for relevant models.


\section{Discussion and Future Work}
\label{sec:conclusions}

We presented an improved algorithm for the construction of MPOs and TTNOs, addressing key limitations in existing methods. We found that our SGE-based algorithm achieves optimal bond dimensions and outperforms the original purely bipartite-graph-based algorithm of \cite{Ren_2020} for a diverse set of Hamiltonians. A mathematical proof of optimality is still outstanding. Furthermore, the added pre-processing step introduces a computational overhead. However, this is acceptable as the TTNO must be created merely once at the simulation's outset. This is insignificant compared to the computational cost of tensor network simulations. Additionally, the symbolic framework used allows for the reuse of a constructed TTNO with different parameters, allowing reuse across multiple simulations.

An additional point of interest is the reliance on predefined tree structures to construct the TTNOs. This leaves room for further exploration into optimal configurations. Advances of this have been made for given quantum states and might be adapted to the construction of TTNOs \cite{Okunishi2023, Hikihara2023, Watanabe2024}. Future work could also consider finding ways of automatic construction for projected entangled pair operators (PEPOs), which are the most general way to represent quantum operators as a tensor network \cite{Cirac2021}. Our algorithm is readily applicable to construct matrix product states or tree tensor network states that are given in a symbolic form. This can be utilized to generate the initial state before being input into a simulation. We also remark that the symbolic nature of our methodology enables gradient computation with respect to the Hamiltonian coefficients.

\begin{acknowledgments}
Thanks to Yaling Ke for the helpful discussion regarding the uses of TTN and the guidance with regard to the HEOM method. The research is part of the Munich Quantum Valley, which is supported by the Bavarian state government with funds from the Hightech Agenda Bayern Plus. Also, financial support for Hazar Çakır through the joint scholarship program of the German Academic Exchange Service (DAAD) and the Turkish Education Foundation (TEV) is gratefully acknowledged.
\end{acknowledgments}


\appendix
\counterwithin*{equation}{section}
\renewcommand\theequation{\thesection\arabic{equation}}
\counterwithin*{figure}{section}
\renewcommand\thefigure{\thesection.\arabic{figure}}

\section{Abelian Symmetries and Quantum Numbers}
\label{appendix:quantum_numbers}

Many quantum Hamiltonians adhere to conservation laws, like global particle or spin conservation. Such abelian symmetries can be used to endow the tensors of the corresponding MPO or TTNO representation with a block sparsity structure \cite{McCulloch2007, Singh2010, Singh2011}. A natural approach to realize this idea in practice is to fix a direction for each tensor leg (inward or outward of the tensor) and to store an ordered list of (integer or half-integer) quantum numbers for each leg. The directions and quantum numbers must be compatible: an outward leg of one tensor can only be contracted with an inward leg of another tensor, and the respective quantum numbers must match. The sparsity structure then follows from the rule that a tensor entry can only be non-zero if the sum of ingoing quantum numbers equals the sum of outgoing quantum numbers. For example, consider the following degree-three tensor $T \in \C^{d_1 \times d_2 \times d_3}$, with axis directions indicated by the arrows:
\[
\begin{tikzpicture}[style={inner sep=0}, decoration={markings, mark=at position 0.5 with {\arrow{>}}}]
\node[draw,circle,thick,minimum size=24] (a) at (0, 0) {$T$};
\node at (0, 1.2) {\footnotesize $i$};
\node at ({1.2*cos(210)}, {1.2*sin(210)}) {\footnotesize $j$};
\node at ({1.2*cos(-30)}, {1.2*sin(-30)}) {\footnotesize $k$};
\draw[postaction={decorate}] (a) --node[right=1mm] {\footnotesize $q_1$} (0, 1);
\draw[postaction={decorate}] (a) -- ({cos(210)}, {sin(210)});
\node at (-0.7, -0.1) {\footnotesize $q_2$};
\draw[postaction={decorate}] ({cos(-30)}, {sin(-30)}) -- (a);
\node at (0.7, -0.1) {\footnotesize $q_3$};
\end{tikzpicture}
\]
The length of each quantum number list matches the respective dimension: $q_1 \in (\frac{1}{2} \Z)^{d_1}$, $q_2 \in (\frac{1}{2} \Z)^{d_2}$, $q_3 \in (\frac{1}{2} \Z)^{d_3}$. In this example, the entry $T_{ijk}$ (for $i = 1, \dots, d_1$, $j = 1, \dots, d_2$, $k = 1, \dots, d_3$) can only be non-zero if
\begin{equation}
q_{1,i} + q_{2,j} = q_{3,k}.
\end{equation}

While symmetries are not the focus of the present work, we want to highlight a simple procedure of including quantum numbers in state diagrams: one associates a single quantum number with each vertex of the state diagram. Since the vertices form the virtual bonds, this assignment specifies the virtual bond quantum numbers. The quantum numbers for the physical axes (tensor legs) are usually known based on the underlying physical model, like $\{\pm\frac{1}{2}\}$ for a spin-$\frac{1}{2}$ chain. The quantum numbers for the vertices are determined by the surrounding local operators. For example, given the local basis $(\ket{\uparrow}, \ket{\downarrow})$ and spin operators $S_{+} = \left(\begin{smallmatrix} 0 & 1 \\ 0 & 0 \end{smallmatrix}\right)$ and $S_{-} = \left(\begin{smallmatrix} 0 & 0 \\ 1 & 0 \end{smallmatrix}\right)$, the term $S_{+}^{(\ell)} S_{-}^{(\ell+1)}$ represented as state diagram with attached quantum numbers (red) reads
\[
\begin{tikzpicture}[
default node/.style={draw, minimum size=0.5cm, thick},
edge node/.style={fill, circle, scale=0.5}
]
\node[default node, draw] (sup) at (-1,0) {$S_{+}^{(\ell)}$};
\node[default node, draw] (sdn) at ( 1,0) {$S_{-}^{(\ell+1)}$};
\node[edge node] (e0) at (-2, 0) {};
\node[edge node] (e1) at ( 0, 0) {};
\node[edge node] (e2) at ( 2, 0) {};
\node at (-2, 0.3) {\small ${\color{red}0}$};
\node at ( 0, 0.3) {\small ${\color{red}1}$};
\node at ( 2, 0.3) {\small ${\color{red}0}$};
\draw (-2.5, 0) -- (e0) -- (sup) -- (e1) -- (sdn) -- (e2) -- ( 2.5, 0);
\end{tikzpicture}
\]
The central vertex has quantum number $1$ since $S_{+}$ flips $\ket{\downarrow}$ to $\ket{\uparrow}$, corresponding to an overall spin change by $1$.

The algorithms described in the following work similarly with quantum numbers attached to the vertices.

\section{Limitations of the Bipartite Graph Algorithm}
\label{app:hamiltonian_example}

\subsection{Example: Non-Optimality in the Bipartite Algorithm}\label{app:minimum_example_bp_not_optimal}
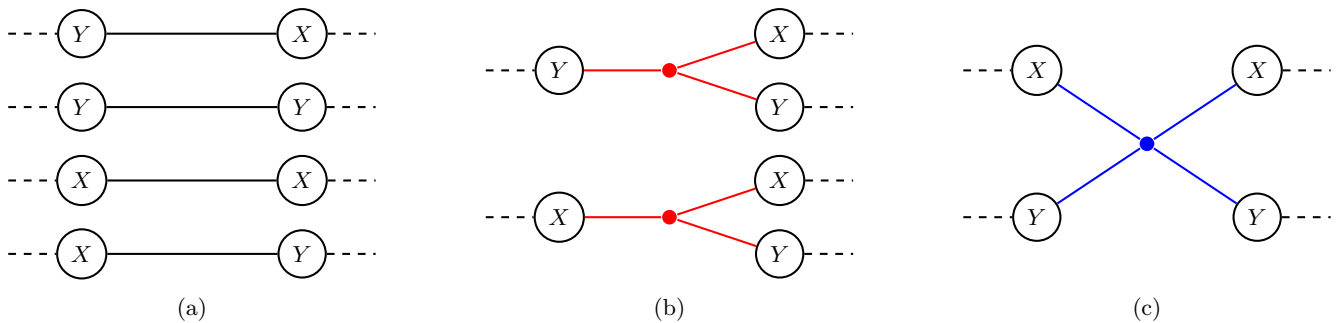
\begin{figure*}[t]
    \centering
    \resizebox{\linewidth}{!}{
    \begin{tikzpicture}

\node at (1.5, -0.75) {(a)};

\node[circle, draw] (X11) at (0, 0) {\(X\)};
\node[circle, draw] (Y21) at (3, 0) {\(Y\)};
\draw (X11) -- (Y21);
\draw[dashed] (-1, 0) -- (X11);
\draw[dashed] (Y21) -- (4, 0);

\node[circle, draw] (X12) at (0, 1) {\(X\)};
\node[circle, draw] (X21) at (3, 1) {\(X\)};
\draw (X12) -- (X21);
\draw[dashed] (-1, 1) -- (X12);
\draw[dashed] (X21) -- (4, 1);

\node[circle, draw] (Y11) at (0, 2) {\(Y\)};
\node[circle, draw] (Y22) at (3, 2) {\(Y\)};
\draw (Y11) -- (Y22);
\draw[dashed] (-1, 2) -- (Y11);
\draw[dashed] (Y22) -- (4, 2);

\node[circle, draw] (Y12) at (0, 3) {\(Y\)};
\node[circle, draw] (X22) at (3, 3) {\(X\)};
\draw (Y12) -- (X22);
\draw[dashed] (-1, 3) -- (Y12);
\draw[dashed] (X22) -- (4, 3);

\begin{scope}[shift={(0,0)}]
    \node at (8, -0.75) {(b)};
    
    \node[circle, draw] (BX) at (6.5, 0.5) {\(X\)};
    \node[circle, draw] (BX1) at (9.5, 1) {\(X\)};
    \node[circle, draw] (BY1) at (9.5, 0) {\(Y\)};
    \node[fill=red, circle, inner sep=2pt] (mid1) at (8, 0.5) {};
    \draw[dashed] (5.5, 0.5) -- (BX);
    \draw[dashed] (BY1.east) -- (10.5, 0);
    \draw[dashed] (BX1.east) -- (10.5, 1);
    \draw[red, thick] (BX) -- (mid1) -- (BY1);
    \draw[red, thick] (mid1) -- (BX1);
    
    \node[circle, draw] (BY) at (6.5, 2.5) {\(Y\)};
    \node[circle, draw] (BX2) at (9.5, 3) {\(X\)};
    \node[circle, draw] (BY2) at (9.5, 2) {\(Y\)};
    \node[fill=red, circle, inner sep=2pt] (mid2) at (8, 2.5) {};
    \draw[dashed] (5.5, 2.5) -- (BY);
    \draw[dashed] (BY2.east) -- (10.5, 2);
    \draw[dashed] (BX2.east) -- (10.5, 3);
    \draw[red, thick] (BY) -- (mid2) -- (BY2);
    \draw[red, thick] (mid2) -- (BX2);
\end{scope}

\begin{scope}[shift={(0,0)}]
    \node at (14.5, -0.75) {(c)};
    \node[circle, draw] (CX1) at (13, 2.5) {\(X\)};
    \node[circle, draw] (CY1) at (13, 0.5) {\(Y\)};
    \node[circle, draw] (CX2) at (16, 2.5) {\(X\)};
    \node[circle, draw] (CY2) at (16, 0.5) {\(Y\)};
    \node[fill=blue, circle, inner sep=2pt] (cmid) at (14.5, 1.5) {};
    \draw[dashed] (12, 2.5) -- (CX1);
    \draw[dashed] (12, 0.5) -- (CY1);
    \draw[dashed] (CX2.east) -- (17, 2.5);
    \draw[dashed] (CY2.east) -- (17, 0.5);
    \draw[blue, thick] (CX1) -- (cmid) -- (CY1);
    \draw[blue, thick] (CX2) -- (cmid) -- (CY2);
\end{scope}

\end{tikzpicture}
    }
    \caption{a) The initial configuration of the chains. b) The result of the bipartite-graph algorithm with virtual dimension $2$. c) Actual optimal solution with a node connected to two operators on both sides.}
    \label{fig:bipartite_fc}
\end{figure*}

Here we consider a minimum example for which the bipartite-graph-based construction algorithm described in Sec.~\ref{sec:bipartite_method} fails to find the optimal result. It is given by the $4$-term Hamiltonian
\begin{align}
H_f &= \sum_{j = 1}^4 \gamma h_j \nonumber\\
    &= \gamma \left( X_1Y_1 + X_1Y_2 + X_2Y_1 + X_2Y_2 \right).
\end{align}
Fig.~\ref{fig:bipartite_fc} compares the initial configuration, the solution proposed by the bipartite algorithm, and the actual optimal solution. The corresponding \( \Gamma_f \)-matrix at the only bond is
\begin{equation}
\Gamma_f = 
\begin{pmatrix}
    \gamma & \gamma \\
    \gamma & \gamma
\end{pmatrix}.
\end{equation}
Thus the optimal way to decompose $\Gamma_f$ is
\begin{equation}
\Gamma_f
= \begin{pmatrix}
    1 \\
    1
\end{pmatrix}
\begin{pmatrix}
    \gamma & \gamma
\end{pmatrix}.
\end{equation}
This is clearly a rank-$1$ decomposition. The bipartite graph algorithm, however, fails to recognize such cases, which require combining both rows and columns rather than selecting them independently. It will, therefore, find a rank-$2$ solution.

\subsection{Symbolic Gaussian Elimination - Examples}\label{app:SGE}
Here we show the SGE with every step for the example given in Sec.~\ref{sec:example_calc}. In each step, the updates are highlighted with red, and we remove the zero rows-columns as a final step in the end.

\begin{align*}
\tiny
\begin{split}
&\begin{pmatrix}
1 & 0 & 0 & 0 & 0 \\
0 & 1 & 0 & 0 & 0\\
0 & 0 & 1 & 0 & 0\\
0 & 0 & 0 & 1 & 0\\
0 & 0 & 0 & 0 & 1
\end{pmatrix}
\begin{pmatrix}
3\gamma_{11} & 3\gamma_{12} & -\gamma_{13} & 0 & 2\gamma_{12} \\
0 & 3\gamma_{22} & 0 & 0 & 2\gamma_{22} \\
0 & 3\gamma_{32} & 0 & 0 & 2\gamma_{32} \\
0 & 0 & 2\gamma_{43} & 2\gamma_{44} & 0 \\
0 & 0 & -\gamma_{43} & -\gamma_{44} & 0
\end{pmatrix}
\begin{pmatrix}
1 & 0 & 0 & 0 & 0 \\
0 & 1 & 0 & 0 & 0\\
0 & 0 & 1 & 0 & 0\\
0 & 0 & 0 & 1 & 0\\
0 & 0 & 0 & 0 & 1
\end{pmatrix}\\ 
&\xrightarrow{} \begin{pmatrix}
1 & 0 & 0 & 0 & 0 \\
0 & 1 & 0 & 0 & 0\\
0 & 0 & 1 & 0 & 0\\
0 & 0 & 0 & 1 & 0\\
0 & 0 & 0 & \textcolor{red}{-\frac{1}{2}} & 1
\end{pmatrix}
\begin{pmatrix}
3\gamma_{11} & 3\gamma_{12} & -\gamma_{13} & 0 & 2\gamma_{12} \\
0 & 3\gamma_{22} & 0 & 0 & 2\gamma_{22} \\
0 & 3\gamma_{32} & 0 & 0 & 2\gamma_{32} \\
0 & 0 & 2\gamma_{43} & 2\gamma_{44} & 0 \\
0 & 0 & \textcolor{red}0 & \textcolor{red}0 & 0
\end{pmatrix}
\begin{pmatrix}
1 & 0 & 0 & 0 & 0 \\
0 & 1 & 0 & 0 & 0\\
0 & 0 & 1 & 0 & 0\\
0 & 0 & 0 & 1 & 0\\
0 & 0 & 0 & 0 & 1
\end{pmatrix} \\
&\xrightarrow{} \begin{pmatrix}
1 & 0 & 0 & 0 & 0 \\
0 & 1 & 0 & 0 & 0\\
0 & 0 & 1 & 0 & 0\\
0 & 0 & 0 & 1 & 0\\
0 & 0 & 0 & -\frac{1}{2} & 1
\end{pmatrix}
\begin{pmatrix}
3\gamma_{11} & 3\gamma_{12} & -\gamma_{13} & 0 & \textcolor{red}0 \\
0 & 3\gamma_{22} & 0 & 0 & \textcolor{red}0 \\
0 & 3\gamma_{32} & 0 & 0 & \textcolor{red}0 \\
0 & 0 & 2\gamma_{43} & 2\gamma_{44} & 0 \\
0 & 0 & 0 & 0 & 0
\end{pmatrix}
\begin{pmatrix}
1 & 0 & 0 & 0 & 0 \\
0 & 1 & 0 & 0 & \textcolor{red}{\frac{2}{3}}\\
0 & 0 & 1 & 0 & 0\\
0 & 0 & 0 & 1 & 0\\
0 & 0 & 0 & 0 & 1
\end{pmatrix} \\
&\xrightarrow{} \begin{pmatrix}
1 & 0 & 0 & 0  \\
0 & 1 & 0 & 0  \\
0 & 0 & 1 & 0  \\
0 & 0 & 0 & 1  \\
0 & 0 & 0 & -\frac{1}{2}  
\end{pmatrix}
\begin{pmatrix}
3\gamma_{11} & 3\gamma_{12} & -\gamma_{13} & 0 \\
0 & 3\gamma_{22} & 0 & 0 \\
0 & 3\gamma_{32} & 0 & 0 \\
0 & 0 & 2\gamma_{43} & 2\gamma_{44}
\end{pmatrix}
\begin{pmatrix}
1 & 0 & 0 & 0 & 0 \\
0 & 1 & 0 & 0 & \frac{2}{3}\\
0 & 0 & 1 & 0 & 0\\
0 & 0 & 0 & 1 & 0
\end{pmatrix}
\end{split}
\end{align*}

Now, we consider a more complex example where the bipartite graph algorithm, as described in Sec.~\ref{sec:bipartite_method}, is unable to find an optimal virtual bond dimension. This given configuration has a rank of $3$, which actually would be written as the sum of three terms. 

\tiny
\begin{align*}
    \Gamma_h &= M_\ell \cdot \Tilde{\Gamma} \cdot M_r  \\
    \begin{pmatrix}
    \gamma_{1} & \gamma_{2} & 0 & 0 \\
    0      & \gamma_{2} & \gamma_{3} & 0 \\
    \gamma_{1} & 0 & 0 & \gamma_{4} \\
    0      & 0      & \gamma_{3} & \gamma_{4}
    \end{pmatrix} &= 
    \begin{pmatrix}
    1 & 0 & 0 & 0 \\
    0 & 1 & 0 & 0 \\
    0 & 0 & 1 & 0 \\
    0 & 0 & 0 & 1
    \end{pmatrix}
    \cdot
    \begin{pmatrix}
    \gamma_{1} & \gamma_{2} & 0 & 0 \\
    0      & \gamma_{2} & \gamma_{3} & 0 \\
    \gamma_{1} & 0 & 0 & \gamma_{4} \\
    0      & 0      & \gamma_{3} & \gamma_{4}
    \end{pmatrix} 
    \cdot
    \begin{pmatrix}
    1 & 0 & 0 & 0 \\
    0 & 1 & 0 & 0 \\
    0 & 0 & 1 & 0 \\
    0 & 0 & 0 & 1
    \end{pmatrix} \\
    \textit{($R_3$ = $R_3$ - $R_1$)}&= 
    \begin{pmatrix}
    1 & 0 & 0 & 0 \\
    0 & 1 & 0 & 0 \\
    1 & 0 & 1 & 0 \\
    0 & 0 & 0 & 1
    \end{pmatrix}
    \cdot
    \begin{pmatrix}
    \gamma_{1} & \gamma_{2} & 0 & 0 \\
    0      & \gamma_{2} & \gamma_{3} & 0 \\
    0 & -\gamma_{2} & 0 & \gamma_{4} \\
    0      & 0      & \gamma_{3} & \gamma_{4}
    \end{pmatrix} 
    \cdot
    \begin{pmatrix}
    1 & 0 & 0 & 0 \\
    0 & 1 & 0 & 0 \\
    0 & 0 & 1 & 0 \\
    0 & 0 & 0 & 1
    \end{pmatrix}\\
    \textit{($R_3$ = $R_3$ + $R_2$)}&= 
    \begin{pmatrix}
    1 & 0 & 0 & 0 \\
    0 & 1 & 0 & 0 \\
    1 & -1 & 1 & 0 \\
    0 & 0 & 0 & 1
    \end{pmatrix}
    \cdot
    \begin{pmatrix}
    \gamma_{1} & \gamma_{2} & 0 & 0 \\
    0      & \gamma_{2} & \gamma_{3} & 0 \\
    0 & 0 & \gamma_{3} & \gamma_{4} \\
    0      & 0      & \gamma_{3} & \gamma_{4}
    \end{pmatrix}
    \cdot
    \begin{pmatrix}
    1 & 0 & 0 & 0 \\
    0 & 1 & 0 & 0 \\
    0 & 0 & 1 & 0 \\
    0 & 0 & 0 & 1
    \end{pmatrix}\\
    \textit{($R_4$ = $R_4$ - $R_3$)}&= 
    \begin{pmatrix}
    1 & 0 & 0 & 0 \\
    0 & 1 & 0 & 0 \\
    1 & -1 & 1 & 0 \\
    0 & 0 & 1 & 1
    \end{pmatrix}
    \cdot
    \begin{pmatrix}
    \gamma_{1} & \gamma_{2} & 0 & 0 \\
    0      & \gamma_{2} & \gamma_{3} & 0 \\
    0 & 0 & \gamma_{3} & \gamma_{4} \\
    0      & 0      & 0 & 0
    \end{pmatrix}
    \cdot
    \begin{pmatrix}
    1 & 0 & 0 & 0 \\
    0 & 1 & 0 & 0 \\
    0 & 0 & 1 & 0 \\
    0 & 0 & 0 & 1
    \end{pmatrix}\\
    \textit{(Zero Row erased)}&= 
    \begin{pmatrix}
    1 & 0 & 0 \\
    0 & 1 & 0 \\
    1 & -1 & 1 \\
    0 & 0 & 1 
    \end{pmatrix}
    \cdot
    \begin{pmatrix}
    \gamma_{1} & \gamma_{2} & 0 & 0 \\
    0      & \gamma_{2} & \gamma_{3} & 0 \\
    0 & 0 & \gamma_{3} & \gamma_{4} 
    \end{pmatrix}
    \cdot
    \begin{pmatrix}
    1 & 0 & 0 & 0 \\
    0 & 1 & 0 & 0 \\
    0 & 0 & 1 & 0 \\
    0 & 0 & 0 & 1
    \end{pmatrix}
\end{align*}

\normalsize

\section{Implementation Details}
\label{appendix:implementation_details}


\begin{figure}
    \centering
    \begin{tikzpicture}[
    regular node/.style={circle, draw, fill=white, inner sep=2pt, minimum size=25pt},
    every path/.style={thick},
    scale=0.8, 
    transform shape
]

\node[regular node] (1) at (0, 0) {1};
\node[regular node] (2) at (0, 1.5) {2};
\node[regular node] (3) at (2, 2.5) {3};
\node[regular node] (4) at (-2, 2.5) {4};
\node[regular node] (5) at (1, -1.5) {5};
\node[regular node] (6) at (3, -1.5) {6};
\node[regular node] (7) at (0, -3) {7};
\node[regular node] (8) at (-2, -2) {8};

\draw[thick, orange] (1) -- (2);
\draw[thick, darkgreen] (2) -- (3);
\draw[thick, darkgreen] (2) -- (4);
\draw[thick, orange] (1) -- (5);
\draw[thick, darkgreen] (5) -- (6);
\draw[thick, darkgreen] (5) -- (7);
\draw[thick, pink] (7) -- (8);

\draw[dashed, red, thick] (3, -1.5) ellipse [x radius=0.5cm, y radius=0.5cm];
\draw[dashed, blue, rounded corners=10pt, thick] 
    ($(8) + (-0.6,-0.5)$) -- ($(7) + (0.1,-0.9)$) -- 
    ($(5) + (0.9,0)$) -- ($(1) + (0.8,0)$) -- 
    ($(2) + (0.5,-0.5)$) -- ($(3) + (0.6,-0.8)$) -- ($(3) + (0.7,0.8)$)-- ($(2) + (0,1.3)$)--
    ($(4) + (-0.6,0.8)$) -- 
    ($(4) + (-0.7,-0.8)$) -- ($(2) + (-0.5,-0.5)$) -- 
    ($(1) + (-0.8,-0.5)$) -- ($(8) + (-0.6,0.5)$) --cycle;
\draw[dashed, red, rounded corners=10pt, thick] 
    ($(8) + (-0.5,-0.4)$) -- ($(7) + (0.1,-0.8)$) -- 
    ($(7) + (0.6,0)$) -- ($(7) + (0.25,0.7)$)  -- ($(8) + (0,0.7)$) --
    ($(8) + (-0.6,0.3)$) -- cycle;

\draw[dashed, violet, rounded corners=10pt, thick] 
    ($(1) + (0.7,0)$) -- 
    ($(2) + (0.4,-0.4)$) -- ($(3) + (0.5,-0.6)$) -- ($(3) + (0.6,0.7)$)-- ($(2) + (0.5,1.1)$)--
    ($(2) + (-0.5,1.1)$)-- ($(4) + (-0.5,0.7)$) -- 
    ($(4) + (-0.6,-0.6)$) -- ($(2) + (-0.4,-0.4)$) -- 
    ($(1) + (-0.65,-0.5)$) -- ($(1) + (0.3,-0.8)$)  -- cycle;

\end{tikzpicture}

    \caption{Each V-subtree (blue), consist of previously calculated U-subtrees (red) and previously calculated V-subtrees (purple).}
    \label{fig:graph_v_detailed}
\end{figure}
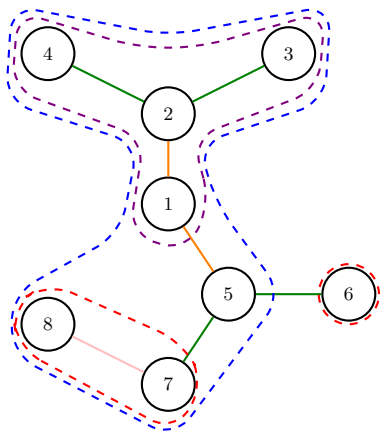

To facilitate comparisons within the tree structure, we use Merkle tree hashing \cite{Merkle1979}. Each node is assigned a hash value, computed based on its internal content and the hash values of its children. These hash values are calculated once during the tree's creation, working from the bottom up. With this approach, comparing two subtrees reduces to comparing their precomputed hash values, significantly reducing the computational burden. Without such a mechanism, comparing large trees would require examining hundreds or thousands of nodes for each comparison.

However, the hash values cannot be directly applied to \( V \)-subtrees, as these do not form proper subtree structures. Instead, \( V \)-subtrees are uniquely identified using the cut-site node and its connections to other subtrees. As illustrated in Figure \ref{fig:graph_v_detailed}, each \( V \)-subtree is composed of either \( U \)-subtrees or previously calculated \( V \)-subtrees. We developed a specialized hash function to incorporate the internal information of the cut-site node and the hash values of its connected subtrees, simplifying comparisons.

The double-traversed BFS enables efficient computation of \( U \)-subtrees and \( V \)-subtrees for earlier depths before the cut, allowing effective reuse of precomputed information. Additionally, the algorithm imposes no restriction on which node to choose as root since the Hamiltonians do not inherently have a predefined root. An arbitrary node can be selected as the root, initiating the algorithm from any node within the tree structure. This flexibility ensures that the processing order does not affect the optimization results but reduces computational complexity. This implementation can be found in the PyTreeNet library \cite{milbradt2024pytreenet, PyTreeNet}.

\section{Derivation of the Super-Hamiltonian}\label{appendix:der_of_sham}

In this appendix we provide further details on the derivation of the effective super-Hamiltonian in \eqref{eq:heom_eff_ham}, given the system Hamiltonian $H_S$ stated in \eqref{eq:system_ham}. As stated in the main text, we want to turn the create the Hamiltonian into a form usable by the HEOM method.

For this we have to approximate the environment $E$ as a collection of bosonic modes. In the case of multiple molecules and a cavity, we approximate the environment with $N_{s}$-many bosonic modes per molecule for the solution and $N_{c}$-many bosonic modes for the electromagnetic cavity environment. For ease of notation we assume $N_{s} = N_{c} = \mathcal{N}$. Now we can use HEOM to determine the dynamics of the system state, given by the density matrix $\rho(t)$. HEOM introduces auxiliary density operators $\rho^{K} (t)$ with the $\mathcal{N} \times (\mathcal{M}+1)$ index matrix $R$. Assuming independent baths for each molecule and the cavity, the HEOM are given by the coupled differential equations
\begin{equation}\label{eq:heom}
\begin{split}
    i & \frac{d}{dt} \rho^{R} = \left[ H_S,\rho^R \right] - i\sum_{\alpha=0}^{\mathcal{M}} \sum_{\beta=1}^{\mathcal{N}} R_{\alpha\beta} \gamma_{\alpha\beta} \rho^R\\
    & + \sum_{\alpha=0}^{\mathcal{M}} \sum_{\beta=1}^{\mathcal{N}} \sqrt{R_{\alpha\beta} + 1} \left( \lambda_\alpha L_\alpha \rho^{R_+^{(\alpha\beta)}} - \lambda_\alpha^* \rho^{R_+^{(\alpha\beta)}} L_\alpha^\dagger \right) \\
    & + \sum_{\alpha=0}^{\mathcal{M}} \sum_{\beta=1}^{\mathcal{N}} \sqrt{R_{\alpha\beta}} \left( \chi_{\alpha\beta} L_\alpha \rho^{R_-^{(\alpha\beta)}} - \chi_{\alpha\beta}^* \rho^{R_-^{(\alpha\beta)}} L_\alpha^\dagger \right),
\end{split}
\end{equation}
where $\gamma_{\alpha\beta}$, $\lambda_\alpha$, and $\chi_{\alpha\beta}$ are complex constants obtained from fitting the actual environment to a finite number of bosonic modes and $L_\alpha$ are the operators coupling the system to the environment acting only on subsystem $\alpha$. The elements of the index matrix $R_\pm^{(\alpha\beta)}$ are given by
\begin{equation}
    \left(R_\pm^{(\alpha\beta)}\right)_{i,j} = \begin{cases}
        R_{i,j} \pm 1 \quad \text{if} \quad (i,j) = (\alpha,\beta) \\
        R_{i,j} \quad \text{else}.
    \end{cases}
\end{equation}
Furthermore, we left out the density matrices' time dependence to shorten the notation. We reinterpret the hierarchy indices as a set of new subsystems \cite{Shi2018, Mangaud2023}. By vectorising the density matrices $\rho^{(K)}$ and introducing the creation operators
\begin{equation}
    b^\dagger_{\alpha\beta} \ket{R} = \sqrt{R_{\alpha\beta} + 1} \ket{R+E_{\alpha\beta}}
\end{equation}
and the annihilation operators
\begin{equation}
    b_{\alpha\beta}\ket{R} = \sqrt{R_{\alpha\beta}} \ket{R-E_{\alpha\beta}}
\end{equation}
we can recast the HEOM \eqref{eq:heom} into an effective Schr\"odinger equation where the Hamiltonian is the super-Hamiltonian $\mathcal{H}$ in Eq.~\eqref{eq:heom_eff_ham}.
\bibliography{references}

\end{document}